\documentclass{pnastwo}

\usepackage{graphicx}
\usepackage{amssymb,amsfonts,amsmath}
\usepackage{enumitem}
\usepackage{rotating}
\usepackage{xcolor}
\usepackage{multirow}

\contributor{Accepted for Computers in Biology and Medicine}
\copyrightyear{2014}
\issuedate{Issue Date}
\volume{Volume}
\issuenumber{Issue Number}

\begin{document}


\title{Automating Fault Tolerance in High-Performance Computational Biological Jobs Using Multi-Agent Approaches} 






\author{Blesson Varghese\affil{1}{University of St Andrews, UK},
Gerard McKee\affil{2}{Baze University, Nigeria}
\and
Vassil Alexandrov\affil{3}{Barcelona Supercomputing Centre, Spain}}

\contributor{Accepted for Computers in Biology and Medicine DOI:10.1016/j.compbiomed.2014.02.005}


\maketitle 

\begin{article}


\begin{abstract}
Background: Large-scale biological jobs on high-performance computing systems require manual intervention if one or more computing cores on which they execute fail. This places not only a cost on the maintenance of the job, but also a cost on the time taken for reinstating the job and the risk of losing data and execution accomplished by the job before it failed. Approaches which can proactively detect computing core failures and take action to relocate the computing core's job onto reliable cores can make a significant step towards automating fault tolerance. \\

Method: This paper describes an experimental investigation into the use of multi-agent approaches for fault tolerance. Two approaches are studied, the first at the job level and the second at the core level. The approaches are investigated for single core failure scenarios that can occur in the execution of parallel reduction algorithms on computer clusters. A third approach is proposed that incorporates multi-agent technology both at the job and core level. Experiments are pursued in the context of genome searching, a popular computational biology application.\\

Result: The key conclusion is that the approaches proposed are feasible for automating fault tolerance in high-performance computing systems with minimal human intervention. In a typical experiment in which the fault tolerance is studied, centralised and decentralised checkpointing approaches on an average add 90\% to the actual time for executing the job. On the other hand, in the same experiment the multi-agent approaches add only 10\% to the overall execution time. 
\end{abstract}








\section{Introduction}

\dropcap{T}he scale of resources and computations required for executing large-scale biological jobs are significantly increasing \cite{01, 02}. With this increase the resultant number of failures while running these jobs will also increase and the time between failures will decrease \cite{03, 03a, 04}. It is not desirable to have to restart a job from the beginning if it has been executing for hours or days or months \cite{04a}. A key challenge in maintaining the seamless (or near seamless) execution of such jobs in the event of failures is addressed under research in fault tolerance \cite{04b, 04c, 04d, 04e}. 

Many jobs rely on fault tolerant approaches that are implemented in the middleware supporting the job (for example \cite{04a, 05, 06, 07}). The conventional fault tolerant mechanism supported by the middleware is checkpointing \cite{07a, 08, 09, 10}, which involves the periodic recording of intermediate states of execution of a job to which execution can be returned if a fault occurs. Such traditional fault tolerant mechanisms, however, are challenged by drawbacks such as single point failures \cite{11}, lack of scalability \cite{11a} and communication overheads \cite{12}, which pose constraints in achieving efficient fault tolerance when applied to high-performance computing systems. Moreover, many of the traditional fault tolerant mechanisms are manual methods and require human administrator intervention for isolating recurring faults. This will place a cost on the time required for maintenance. 

Self-managing or automated fault tolerant approaches are therefore desirable, and the objective of the research reported in this paper is the development of such approaches. If a failure is likely to occur on a computing core on which a job is being executed, then it is necessary to be able to move (migrate) the job onto a reliable core \cite{12a}. Such mechanisms are not readily available. At the heart of this concept is mobility, and a technique that can be employed to achieve this is using multi-agent technologies \cite{13}.

Two approaches are proposed and implemented as the means of achieving both the computation in the job and self-managing fault tolerance; firstly, an approach incorporating agent intelligence, and secondly, an approach incorporating core intelligence. In the first approach, automated fault tolerance is achieved by a collection of agents which can freely traverse on a network of computing cores. Each agent carries a portion of the job (or sub-job) to be executed on a computing core in the form of a payload. Fault tolerance in this context can be achieved since an agent can move on the network of cores, effectively moving a sub-job from one computing core which may fail onto another reliable core. 

In the second approach, automated fault tolerance is achieved by considering the computing cores to be an intelligent network of cores. Sub-jobs are scheduled onto the cores, and the cores can move processes executed on them across the network of cores. Fault tolerance in this context can be achieved since a core can migrate a process executing on it onto another core. 

A third approach is proposed which combines both agent and core intelligence under a single umbrella. In this approach, a collection of agents freely traverse on a network of virtual cores which are an abstraction of the actual hardware cores. The agents carry the sub-jobs as a payload and situate themselves on the virtual cores. Fault tolerance is achieved either by an agent moving off one core onto another core or the core moving an agent onto another core when a fault is predicted. Rules are considered to decide whether an agent or a core should initiate the move.  

Automated fault tolerance can be beneficial in areas such as molecular dynamics \cite{21, 22, 23, 26}. Typical molecular dynamics simulations explore the properties of molecules in gaseous, liquid and solid states. For example, the motion of molecules over a time period can be computed by employing Newton's equations if the molecules are treated as point masses. These simulations require large numbers of computing cores that run sub-jobs of the simulation which communicate with each other for hours, days and even months. It is not desirable to restart an entire simulation or to loose any data from previous numerical computations when a failure occurs. Conventional methods like periodic checkpointing keep track of the state of the sub-jobs executed on the cores, and helps in restarting a job from the last checkpoint. However, overzealous periodic checkpointing over a prolonged period of time has large overheads and contributes to the slowdown of the entire simulation \cite{30}. Additionally, mechanisms will be required to store and handle large data produced by the checkpointing strategy. Further, how wide the failure can impact the simulation is not considered in checkpointing. For example, the entire simulation is taken back to a previous state irrespective of whether the sub-jobs running on a core depend or do not depend on other sub-jobs. 

One potential solution to mitigate the drawbacks of checkpointing is to proactively probe the core for failures. If a core is likely to fail, then the sub-job executing on the core is migrated automatically onto another core that is less likely to fail. This paper proposes and experimentally evaluates multi-agent approaches to realising this automation. Genome searching is considered as an example for implementing the multi-agent approaches. The results indicate the feasibility of the multi-agent approaches; they require only one-fifth the time compared to that required by manual approaches. 

The remainder of this paper is organised as follows. The Methods section presents the three approaches proposed for automated fault tolerance. The Results section highlights the experimental study and the results obtained from it. The Discussion section presents a discussion on the three approaches for automating fault tolerance. The Conclusions section summarises the key results from this study. 


\section{Methods}
\label{methods}

Three approaches to automate fault tolerance are presented in this section. The first approach incorporates agent intelligence, the second approach incorporates core intelligence, and in the third a hybrid of both agent and core intelligence is incorporated. 

\subsection{Approach 1: Fault Tolerance incorporating Agent Intelligence}

A job, $J$, which needs to be executed on a large-scale system is decomposed into a set of sub-jobs $J_{1}, J_{2} \cdots J_{n}$. Each sub-job $J_{1}, J_{2} \cdots J_{n}$ is mapped onto agents $A_{1}, A_{2} \cdots A_{n}$ that carry the sub-jobs as payloads onto the cores, $C_{1}, C_{2} \cdots C_{n}$ as shown in Figure 1. The agents and the sub-job are independent of each other; in other words, an agent acts as a wrapper around a sub-job to situate the sub-job on a core. 

\begin{figure} 
	\centering
	\includegraphics[width=0.5\textwidth]{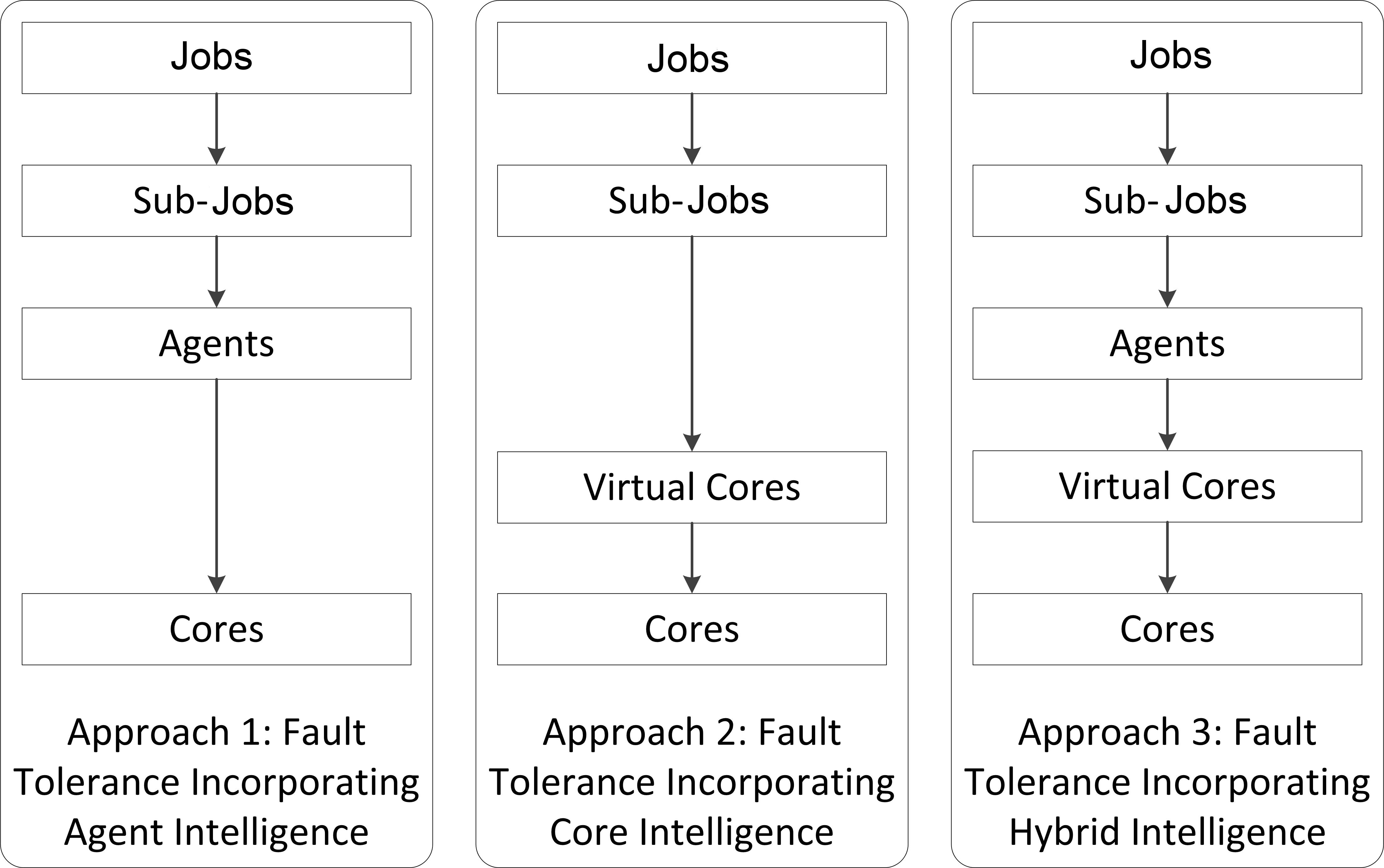}
	\caption{The job, sub-jobs, agents, virtual cores and computing cores in the two approaches proposed for automated fault tolerance}
	\label{figure1}
\end{figure}

There are three computational requirements of the agent to achieve successful execution of the job: (a) the agent needs to know the overall job, $J$, that needs to be achieved, (b) the agent needs to access data required by the sub-job it is carrying and (c) the agent needs to know the operation that the sub-job needs to perform on the data. The agents then displace across the cores to compute the sub-jobs. 

Intelligence of an agent can be useful in at least four important ways for achieving fault tolerance while a sub-job is executed. Firstly, an agent knows the landscape in which it is located. Knowledge of the landscape is threefold which includes (a) the knowledge of the computing core on which the agent is located, (b) knowledge of other computing cores in the vicinity of the agent and (c) knowledge of agents located in the vicinity. Secondly, an agent identifies a location to situate within the landscape. This is possible by gathering information from the vicinity using probing processes and is required when the computing core on which the agent is located is anticipated to fail. Thirdly, an agent predicts failures that are likely to impair its functioning. The prediction of failures (for example, due to the failure of the computing core) is along similar lines to proactive fault tolerance. Fourthly, an agent is mobile within the landscape. If the agent predicts a failure then the agent can relocate onto another computing core thereby moving off the job from the core anticipated to fail (refer Figure 2). 

\begin{figure}
	\centering
	\includegraphics[width=0.4\textwidth]{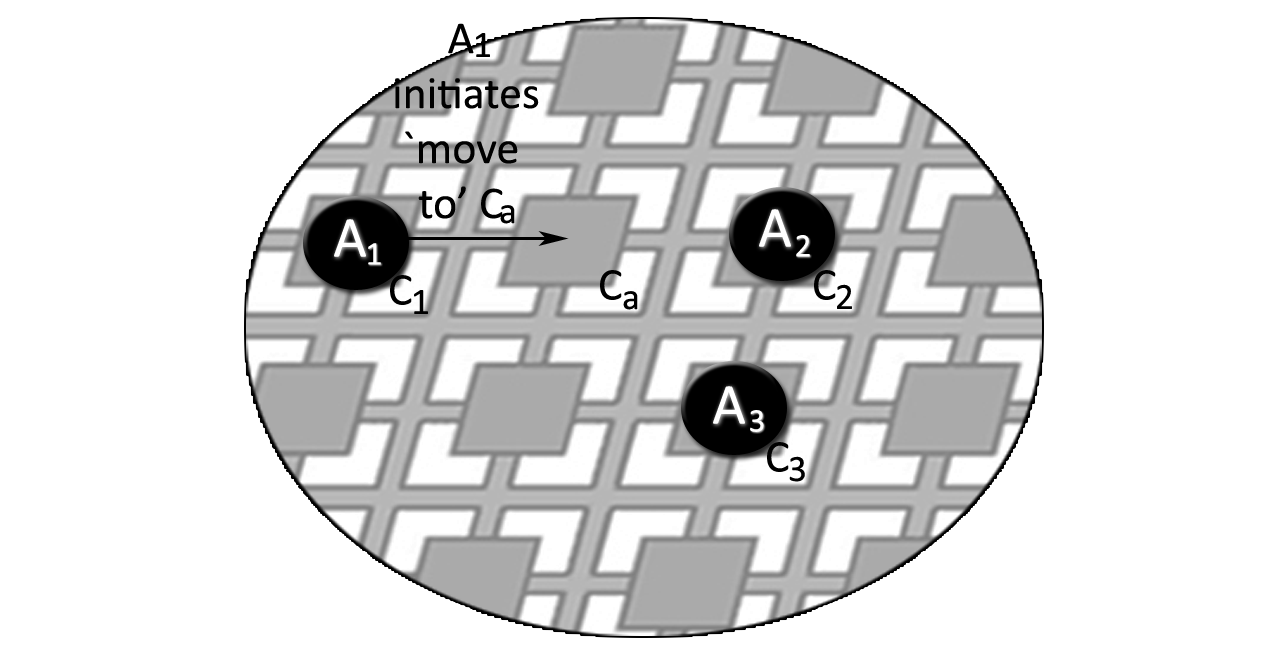}
	\caption{{Agent-Core interaction in Approach 1.} Agents $A_{1}, A_{2}$ and $A_{3}$ are situated on cores $C_{1}, C_{2}$ and $C_{3}$ respectively. A failure is predicted on core $C_{1}$. The agent $A_{1}$ moves onto core $C_{a}$.}
	\label{figure2}
\end{figure}

The intelligence of agents is incorporated within the following sequence of steps that describes an approach for fault tolerance: \vspace{6pt}

\noindent\rule{0.5\textwidth}{0.4pt}
\textit{Agent Intelligence Based Fault Tolerance}\\
\noindent\rule{0.5\textwidth}{0.4pt}

\begin{enumerate}[leftmargin = 1.5cm]
\item[Step 1:] Decompose a job, $J$, to be executed on the landscape into sub-jobs, $J_{1}, J_{2} \cdots J_{n}$
\item[Step 2:] Each sub-job provided as a payload to agents, $A_{1}, A_{2} \cdots A_{n}$
\item[Step 3:] Agents carry jobs onto computing cores, $C_{1}, C_{2} \cdots C_{n}$
\item[Step 4:] For each agent, $A_{i}$ located on computing core $C_{i}$, where $i = 1$ to $n$

\begin{enumerate}[leftmargin = 1.5cm]
\item[Step 4.1:] Periodically probe the computing core $C_{i}$
\item[Step 4.2:] if $C_{i}$ predicted to fail, then	
\end{enumerate}

\begin{enumerate}[leftmargin = 3.3cm]
\item[Step 4.2.1:] Agent, $A_{i}$ moves onto an adjacent computing core, $C_{a}$
\item[Step 4.2.2:] Notify dependent agents
\item[Step 4.2.3:] Agent $A_{i}$ establishes dependencies
\end{enumerate}

\item[Step 5:]Collate execution results from sub-jobs
\end{enumerate}
\noindent\rule{0.5\textwidth}{0.4pt}

\subsubsection{Agent Intelligence Failure Scenario}

A failure scenario is considered for the agent intelligence based fault tolerance concept. In this scenario, while a job is executed on a computing core that is anticipated to fail any adjacent core onto which the job needs to be reallocated can also fail. The communication sequence shown in Figure 3 is as follows. The hardware probing process on the core anticipating failure, $C_{PF}$ notifies the failure prediction to the agent process, $P_{PF}$, situated on it. Since the failure of a core adjacent to the core predicted to fail is possible it is necessary that the predictions of the hardware probing processes on the adjacent cores be requested. Once the predictions are gathered, the agent process, $P_{PF}$, creates a new process on an adjacent core and transfers data it was using onto the newly created process. Then the input dependent (${P_{ID}}_{1} \cdots {P_{ID}}_{n}$) and output dependent (${P_{OD}}_{1} \cdots {P_{OD}}_{n}$) processes are notified. The agent process on $C_{PF}$ is terminated thereafter. The new agent process on the adjacent core establishes dependencies with the input and output dependent processes. 

\begin{figure*} 
	\centering
	\includegraphics[width=0.9\textwidth]{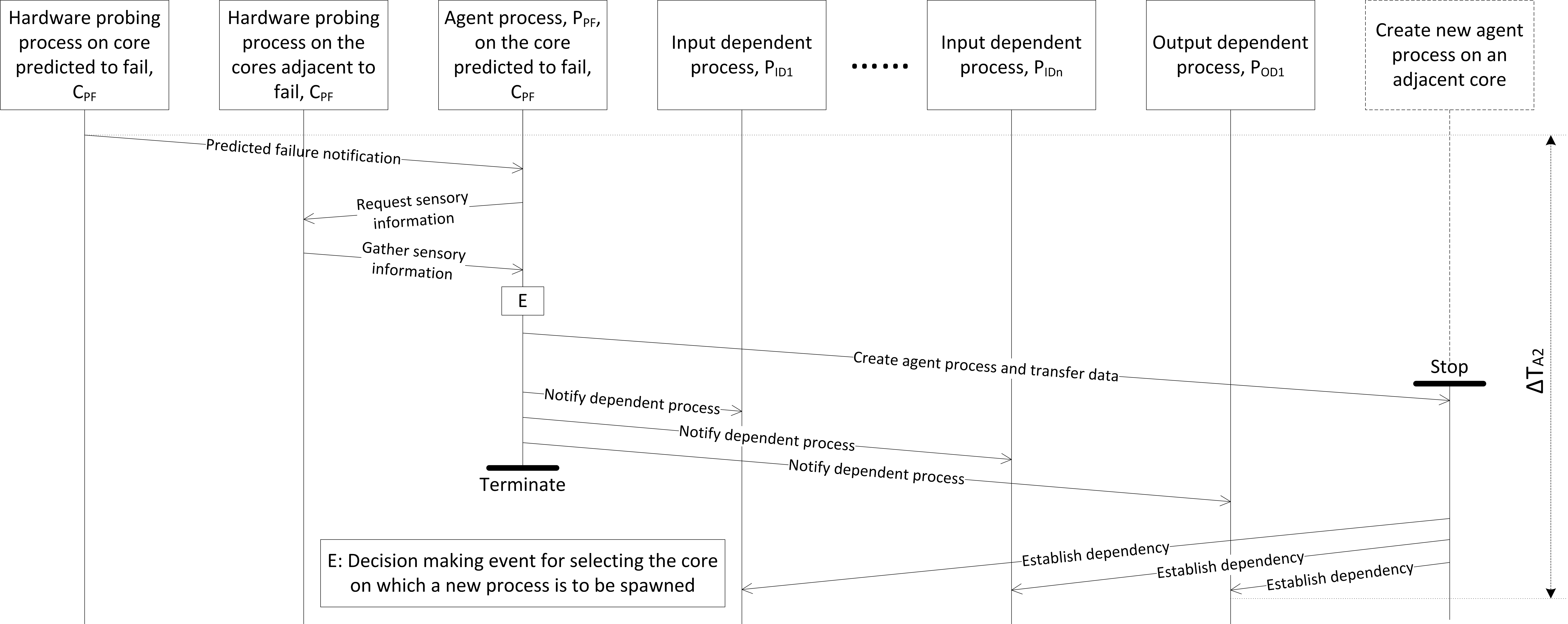}
	\caption{Communication sequence in the failure scenario of agent intelligence based fault tolerance}
	\label{figure3}
\end{figure*}

\subsection{Approach 2: Fault Tolerance incorporating Core Intelligence}

A job, $J$, which needs to be executed on a large-scale system is decomposed into a set of sub-jobs $J_{1}, J_{2} \cdots J_{n}$. Each sub-job $J_{1}, J_{2} \cdots J_{n}$ is mapped onto the virtual cores, $VC_{1}, VC_{2} \cdots VC_{n}$, an abstraction over $C_{1}, C_{2} \cdots C_{n}$ respectively as shown in Figure 4. The cores referred to in this approach are virtual cores which are an abstraction over the hardware computing cores. The virtual cores are a logical representation and may incorporate rules to achieve intelligent behaviour. 

\begin{figure} 
	\centering
	\includegraphics[width=0.4\textwidth]{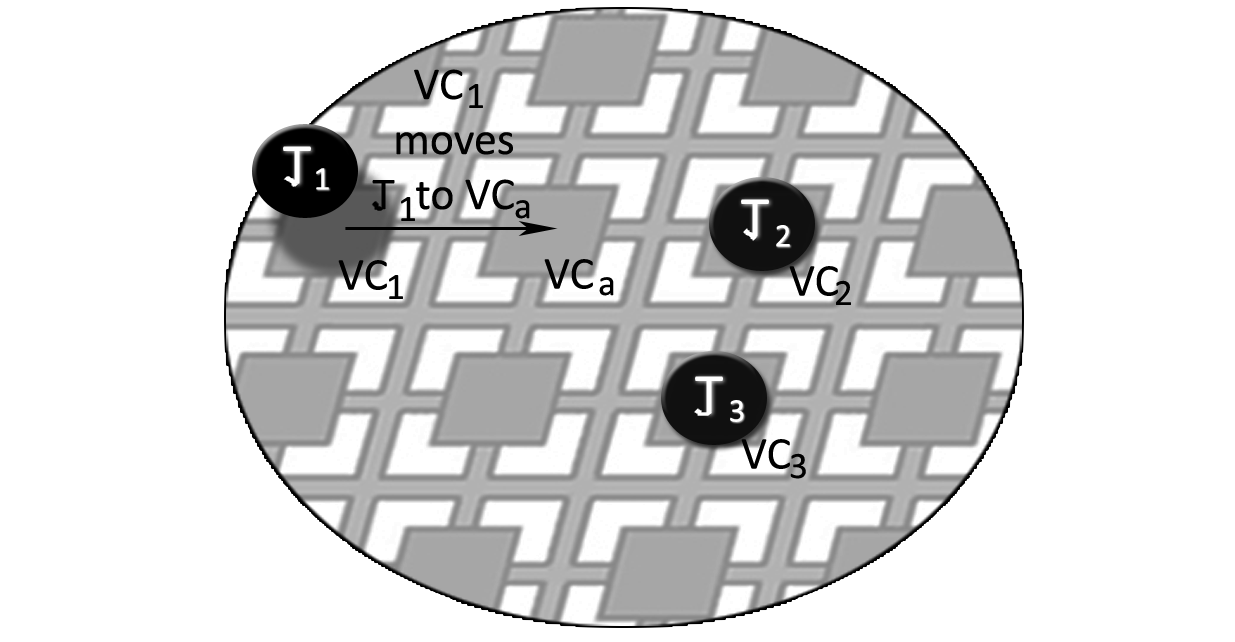}
	\caption{{Job-Virtual Core interaction in Approach 2.} Jobs $J_{1}, J_{2}$ and $J_{3}$ are situated on virtual cores $VC_{1}, VC_{2}$ and $VC_{3}$ respectively. A failure is predicted on core $C_{1}$ and $VC_{1}$ moves the job $J_{1}$ onto virtual core $VC_{a}$.}
	\label{figure4}
\end{figure}

Intelligence of a core is useful in a number of ways for achieving fault tolerance. Firstly, a core updates knowledge of its surrounding by monitoring adjacent neighbours. Independent of what the cores are executing, the cores can monitor each other. Each core can ask the question `are you alive?' to its neighbours and gain information. Secondly, a core periodically updates information of its surrounding. This is useful for the core to know which neighbouring cores can execute a job if it fails. Thirdly, a core periodically monitors itself using a hardware probing process and predicts if a failure is likely to occur on it. Fourthly, a core can move a job executing on it onto an adjacent core if a failure is expected and adjust to failure as shown in Figure 4. Once a job has relocated all data dependencies will need to be re-established.

The following sequence of steps describe an approach for fault tolerance incorporating core intelligence: \vspace{6pt}

\noindent\rule{0.5\textwidth}{0.4pt}
\textit{Core Intelligence Based Fault Tolerance}\\
\noindent\rule{0.5\textwidth}{0.4pt}

\begin{enumerate}[leftmargin = 1.5cm]
\item[Step 1:] Decompose a job, $J$, to be executed on the landscape into sub-jobs, $J_{1}, J_{2} \cdots J_{n}$
\item[Step 2:] Each sub-job allocated to cores, $VC_{1}, VC_{2} \cdots VC_{n}$
\item[Step 3:] For each core, $VC_{i}$, where $i = 1$ to $n$ until sub-job $J_{i}$ completes execution

\begin{enumerate}[leftmargin = 1.5cm]
\item[Step 3.1:] Periodically probe the computing core $C_{i}$
\item[Step 3.2:] if $C_{i}$ predicted to fail, then
\end{enumerate}

\begin{enumerate}[leftmargin = 3.3cm]
\item[Step 3.2.1:] Migrate sub-job $J_{i}$ on $VC_{i}$ onto an adjacent computing core, $VC_{a}$
\end{enumerate}

\item[Step 4:]Collate execution results from sub-jobs
\end{enumerate}
\noindent\rule{0.5\textwidth}{0.4pt}

\subsubsection{Core Intelligence Failure Scenario}

Figure 5 shows the communication sequence of the core failure scenario considered for the core intelligence based fault tolerance concept. The hardware probing process on the core predicted to fail, $C_{PF}$ notifies a predicted failure to the core. The job executed on $VC_{PF}$ is then migrated onto an adjacent core $VC_{1} \cdots VC_{n}$ once a decision based on failure predictions are received from the hardware probing processes of adjacent cores.

\subsection{Approach 3: Hybrid Fault Tolerance}

The hybrid approach acts as an umbrella bringing together the concepts of agent intelligence and core intelligence. The key concept of the hybrid approach lies in the mobility of the agents on the cores and the cores collectively executing a job. Decision-making is required in this approach for choosing between the agent intelligence and core intelligence approaches when a failure is expected.

A job, $J$, which needs to be executed on a large-scale system is decomposed into a set of sub-jobs $J_{1}, J_{2} \cdots J_{n}$. Each sub-job $J_{1}, J_{2} \cdots J_{n}$ is mapped onto agents $A_{1}, A_{2} \cdots A_{n}$ that carry the sub-jobs as payloads onto the virtual cores, $VC_{1}, VC_{2} \cdots VC_{n}$ which are an abstraction over $C_{1}, C_{2} \cdots C_{n}$ respectively as shown in Figure 1.

The following sequence of steps describe the hybrid approach for fault tolerance incorporating both agent and core intelligence: \vspace{6pt}

\noindent\rule{0.5\textwidth}{0.4pt}
\textit{Hybrid Intelligence Based Fault Tolerance}\\
\noindent\rule{0.5\textwidth}{0.4pt}

\begin{enumerate}[leftmargin = 1.5cm]
\item[Step 1:] Decompose a job, $J$, to be executed on the landscape into sub-jobs, $J_{1}, J_{2} \cdots J_{n}$
\item[Step 2:] Each sub-job provided as a payload to agents, $A_{1}, A_{2} \cdots A_{n}$
\item[Step 3:] Agents carry jobs onto virtual cores, $VC_{1}, VC_{2} \cdots VC_{n}$
\item[Step 4:] For each agent, $A_{i}$ located on virtual core $VC_{i}$, where $i = 1$ to $n$

\begin{enumerate}[leftmargin = 1.5cm]
\item[Step 4.1:] Periodically probe the computing core $C_{i}$
\item[Step 4.2:] if $C_{i}$ predicted to fail, then
\end{enumerate}

\begin{enumerate}[leftmargin = 3.3cm]
\item[Step 4.2.1:] if `Agent Intelligence' is a suitable mechanism, then
\end{enumerate}

\begin{enumerate}[leftmargin = 5cm]
\item[Step 4.2.1.1:] Agent, $A_{i}$, moves onto an adjacent computing core, $VC_{a}$
\item[Step 4.2.1.2:] Notify dependent agents
\item[Step 4.2.1.3:] Agent $A_{i}$ establishes dependencies
\end{enumerate}

\begin{enumerate}[leftmargin = 3.3cm]
\item[Step 4.2.2:] else if `Core Intelligence' is a suitable mechanism, then
\end{enumerate}

\begin{enumerate}[leftmargin = 5cm]
\item[Step 4.2.2.1:] Core $VC_{i}$ migrates agent, $A_{i}$ onto an adjacent computing core, $VC_{a}$
\end{enumerate}

\item[Step 5:]Collate execution results from sub-jobs
\end{enumerate}
\noindent\rule{0.5\textwidth}{0.4pt}
\vspace{12pt}

\begin{figure}[t]
	\centering
	\includegraphics[width=0.5\textwidth]{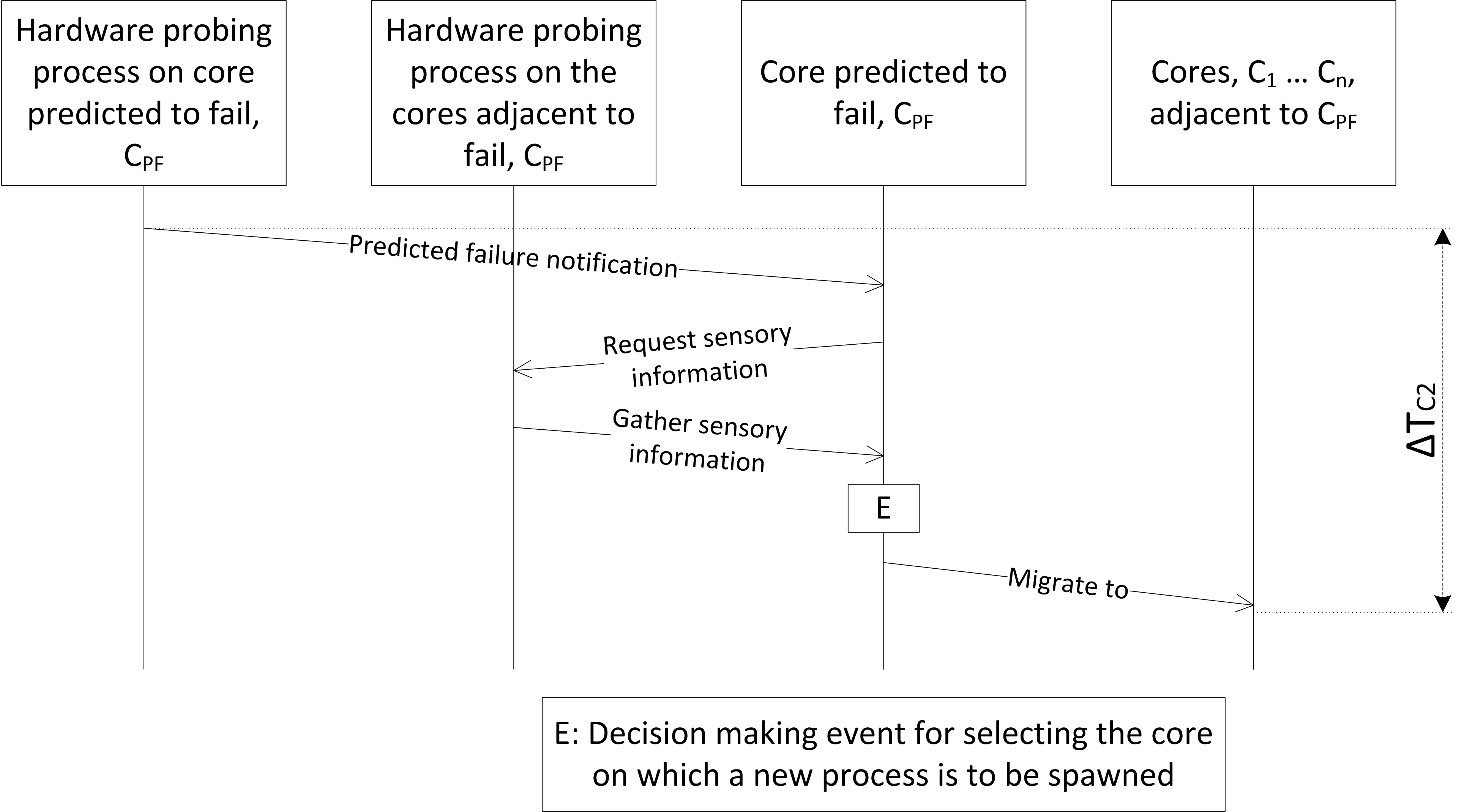}
	\caption{Communication sequence in core intelligence based fault tolerance}
	\label{figure5}
\end{figure}

When a core failure is anticipated both an agent and a core can make decisions which can lead to a conflict. For example, an agent can attempt to move onto an adjacent core while the core on which it is executing would like to migrate it to an alternative adjacent core. Therefore, an agent and the core on which it is located need to negotiate before either of them initiate a response to move (see Figure 6). The rules for the negotiation between the agent and the core in this case are proposed from the experimental results presented in this paper (presented in the Decision Making Rules sub-section).

\begin{figure} 
	\centering
	\includegraphics[width=0.4\textwidth]{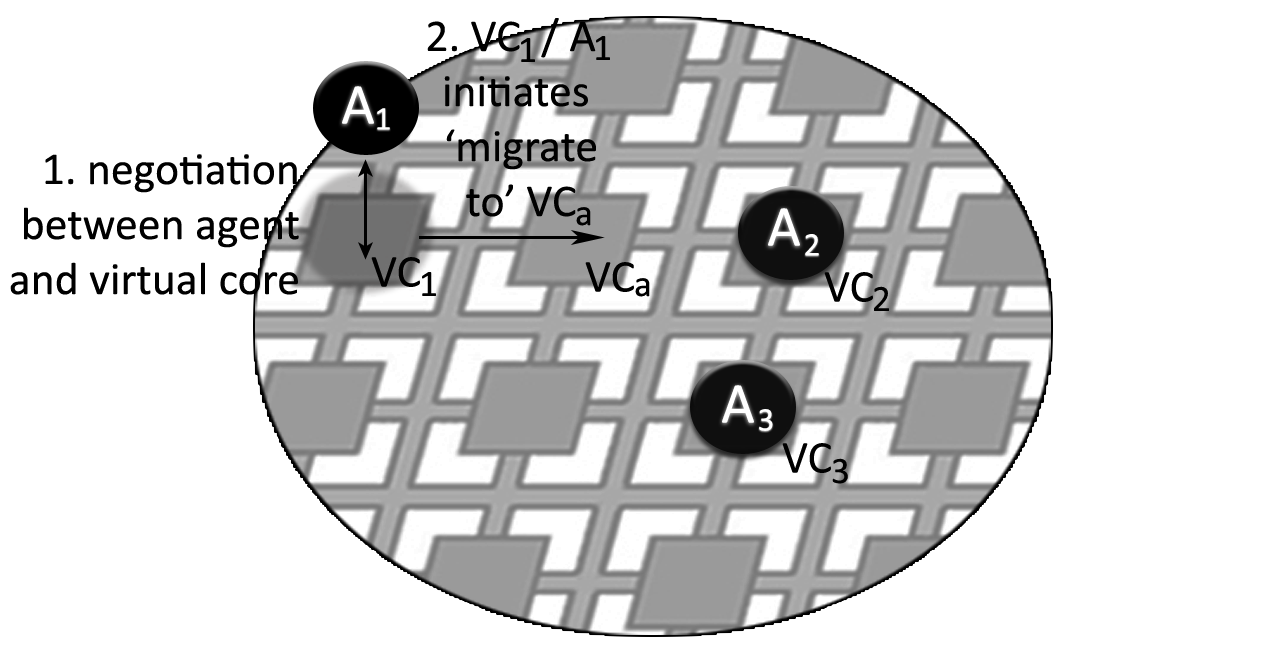}
	\caption{Conflict negotiation and resolution in Approach 3. Agents $A_{1}, A_{2}$ and $A_{3}$ are situated on virtual cores $VC_{1}, VC_{2}$ and $VC_{3}$ which are mapped onto computing cores $C_{1}, C_{2}$ and $C_{3}$ respectively. A failure is predicted on core $C_{1}$. The agent $A_{1}$ and $VC_{1}$ negotiate to decide who moves the sub-job onto core $VC_{a}$.}
	\label{figure6}
\end{figure}


\section{Results}
\label{results}

In this section, the experimental platform is considered followed by the experimental studies and the results obtained from experiments. 

\subsection{Platform}

Four computer clusters were used for the experiments reported in this paper. The first was a cluster available at the Centre for Advanced Computing and Emerging Technologies (ACET), University of Reading, UK. Thirty three compute nodes connected through Gigabit Ethernet were available, each with Pentium IV processors and 512 MB-2 GB RAM. The remaining three clusters are compute resources, namely Brasdor, Glooscap and Placentia, all provided by The Atlantic Computational Excellence Network (ACEnet) \cite{60}, Canada. Brasdor comprises 306 compute nodes connected through Gigabit Ethernet, with 932 cores and 1-2 GB RAM. Glosscap comprises 97 nodes connected through Infiniband, with 852 cores and 1-8 GB RAM. Placentia comprises 338 compute nodes connected through Infiniband, with 3740 cores and 2-16 GB RAM.

The cluster implementations in this paper are based on the Message Passing Interface (MPI). The first approach, incorporating agent intelligence, is implemented using Open MPI \cite{61}, an open source implementation of MPI 2.0. The dynamic process model which supports dynamic process creation and management facilitates control over an executing process. This feature is useful for implementing the first approach. The MPI functions useful in the implementation are (i) MPI\_COMM\_SPAWN which creates a new MPI process and establishes communication with an existing MPI application and (ii) MPI\_COMM\_ACCEPT and MPI\_COMM\_CONNECT which establishes communication between two independent processes. 

The second approach, incorporating core intelligence, is implemented using Adaptive MPI (AMPI) \cite{62}, developed over Charm++ \cite{63}, a C++ based parallel programming language. The aim of AMPI is to achieve dynamic load balancing by migrating objects over virtual cores and thereby facilitating control over cores. Core intelligence harnesses this potential of AMPI to migrate a job from a core onto another core. A strategy to migrate a job using the concepts of processor virtualisation and dynamic job migration in AMPI and Charm++ is reported in \cite{64}. 

\subsection{Experimental Studies}

Parallel reduction algorithms \cite{65, 66} which implement the bottom-up approach (i.e., data flows from the leaves to the root) are employed for the experiments. These algorithms are of interest for three reasons. Firstly, the algorithm is used in a large number of scientific applications including computational biological applications in which optimizations are performed (for example, bootstrapping). Incorporating self-managing fault tolerant approaches can make these algorithms more robust and reliable \cite{67}. Secondly, the algorithm lends itself to be easily decomposed into a set of sub-jobs. Each sub-job can then be mapped onto a computing core either by providing the sub-job as a payload to an agent in the first approach or by providing the job onto a virtual core incorporating intelligent rules. Thirdly, the execution of a parallel reduction algorithm stalls and produces incorrect solutions if a core fails. Therefore, parallel reduction algorithms can benefit from local fault-tolerant techniques.

Figure 7 is an exemplar of a parallel reduction algorithm. In the experiments reported in this paper, a generic parallel summation algorithm with three sets of input is employed. Firstly, $I_{(1,1)}$, $I_{(1,2)}$ $\cdots$ $I_{(1,x)}$, secondly, $I_{(2,1)}$, $I_{(2,2)}$ $\cdots$ $I_{(2,y)}$, and thirdly, $I_{(3,1)}$ $\cdots$ $I_{(3,z)}$. The first level nodes which receive the three sets of input comprise three set of nodes. Firstly, ${N_{1}}_{(1,1)}$, ${N_{1}}_{(1,2)}$ $\cdots$ ${N_{1}}_{(1,x)}$, secondly, ${N_{1}}_{(2,1)}$, ${N_{1}}_{(2,2)}$ $\cdots$ ${N_{1}}_{(2,y)}$, and thirdly, ${N_{1}}_{(3,1)}$, ${N_{1}}_{(3,2)}$ $\cdots$ ${N_{1}}_{(3,z)}$. The next level of nodes, ${N_{2}}_{(1,1)}$, ${N_{2}}_{(2,1)}$ and ${N_{3}}_{(3,1)}$ receive inputs from the first level nodes. The resultant from the second level nodes is fed in to the third level node ${N_{3}}_{(1,1)}$. The nodes reduce the input through the output using the parallel summation operator ($\oplus$).

\begin{figure} 
	\centering
	\includegraphics[width=0.45\textwidth]{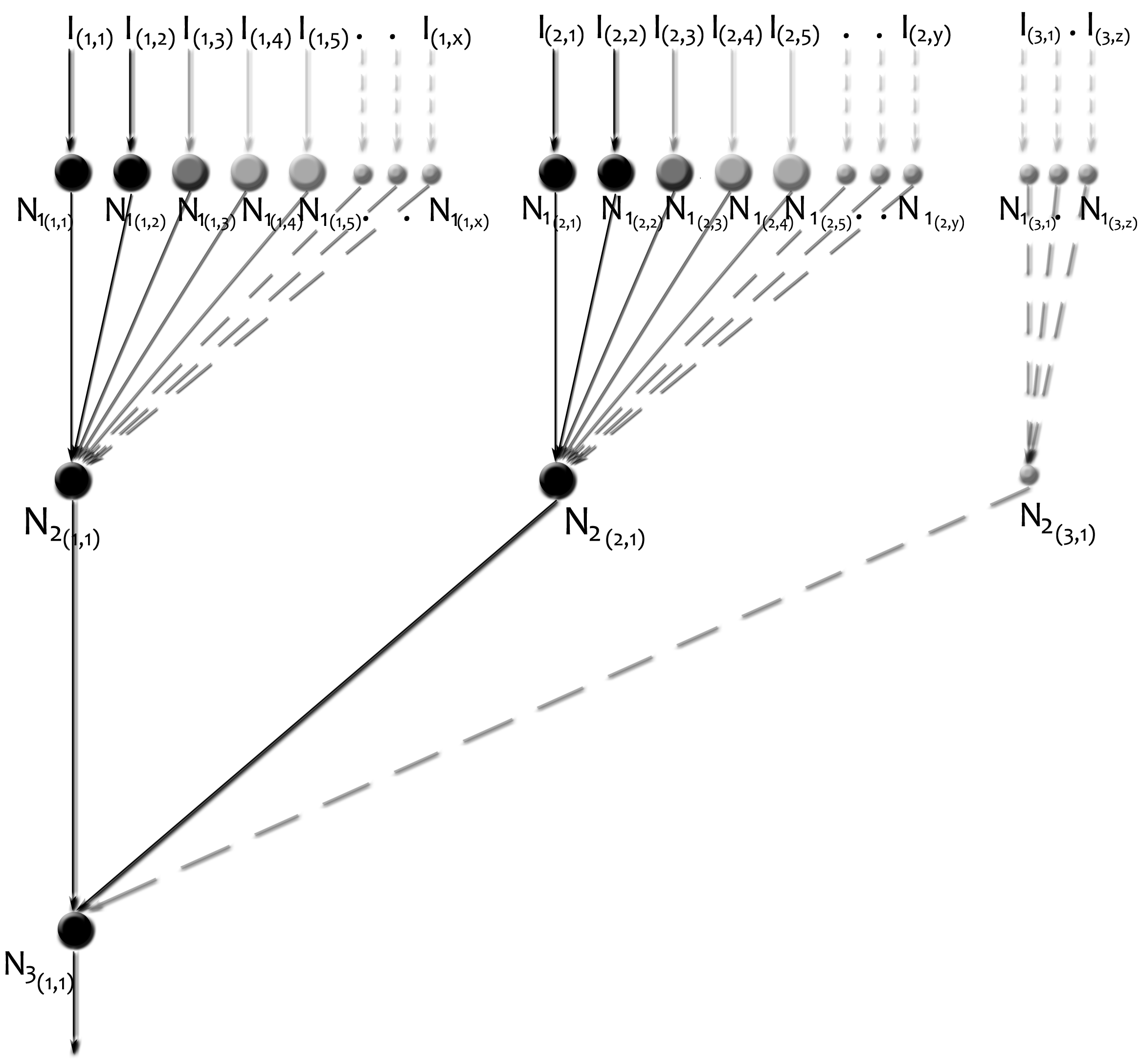}	
	\caption{Generic parallel summation algorithm. The inputs are denoted by $I$ and the three levels of nodes are denoted by $N$. The inputs are passed to the nodes $N_{1}$ which are then reduced and passed to nodes $N_{2}$ and finally onto $N_{3}$ for the output.}
	\label{figure7}
\end{figure}

The parallel summation algorithm can benefit from the inclusion of fault tolerant strategies. The job, $J$, in this case is summation, and the sub-jobs, $J_{1}, J_{2} \cdots J_{n}$ are also summations. In the first fault tolerant approach, incorporating mobile agent intelligence, the data to be summed along with the summation operator is provided to the agent. The agents locate on the computing cores and continuously probe the core for anticipating failures. If an agent is notified of a failure, then it moves off onto another computing core in the vicinity, thereby not stalling the execution towards achieving the summation job. In the second fault tolerant approach, incorporating core intelligence, the sub-job comprising the data to be summed along with the summation operator is located on the virtual core. When the core anticipates a failure, it migrates the sub-job onto another core. 

A failure scenario is considered for experimentally evaluating the fault tolerance strategies. In the scenario, when a core failure is anticipated the sub-job executing on it is relocated onto an adjacent core. Of course this adjacent core may also fail. Therefore, information is also gathered from adjacent cores as to whether they are likely to fail or not. This information is gathered by the agent in the agent-based approach and the virtual core in the core-based approach and used to determine which adjacent core the sub-job needs to be moved to. This failure scenario is adapted to the two strategies giving respectively the agent intelligence failure scenario and the core intelligence failure scenario (described in the Methods section). 

\subsection{Experimental Results}

Figures 8 through 13 are a collection of graphs plotted using the parallel reduction algorithm as a case study for both the first (agent intelligence - Figure 8, Figure 10 and Figure 12) and second (core intelligence - Figure 9, Figure 11 and Figure 13) fault tolerant approaches. Each graph comprises four plots, the first representing the ACET cluster and the other three representing the three ACEnet clusters. The graphs are also distinguished based on the following three factors that can affect the performance of the two approaches:

\begin{figure}
	\centering
	\includegraphics[width=0.49\textwidth]{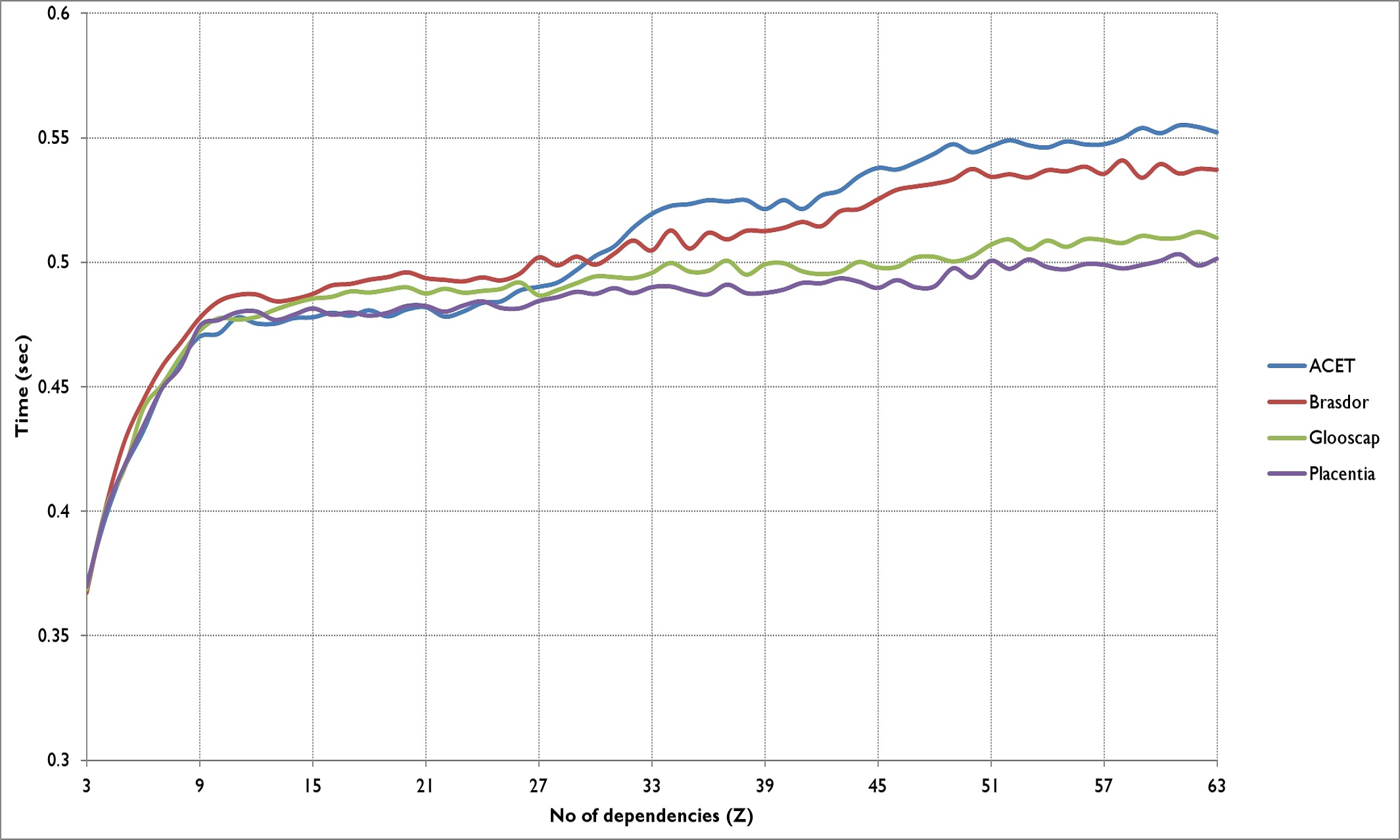}
	\caption{No. of dependencies vs time taken for reinstating execution after failure in the agent intelligent approach}
	\label{figure8}
\end{figure}

\begin{figure}
	\centering
	\includegraphics[width=0.49\textwidth]{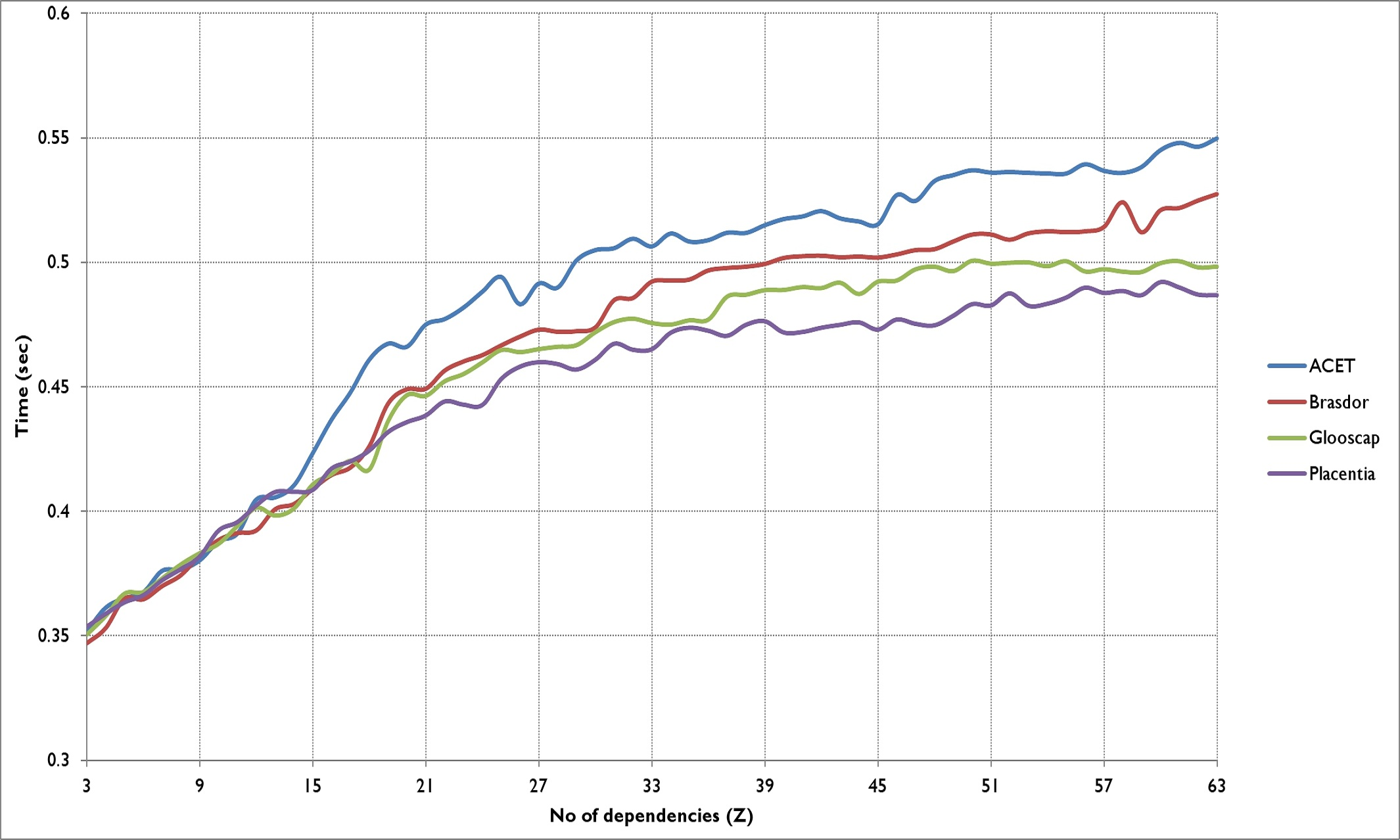}
	\caption{No. of dependencies vs time taken for reinstating execution after failure in the core intelligent approach}
	\label{figure9}
\end{figure}

\begin{figure}
	\centering
	\includegraphics[width=0.49\textwidth]{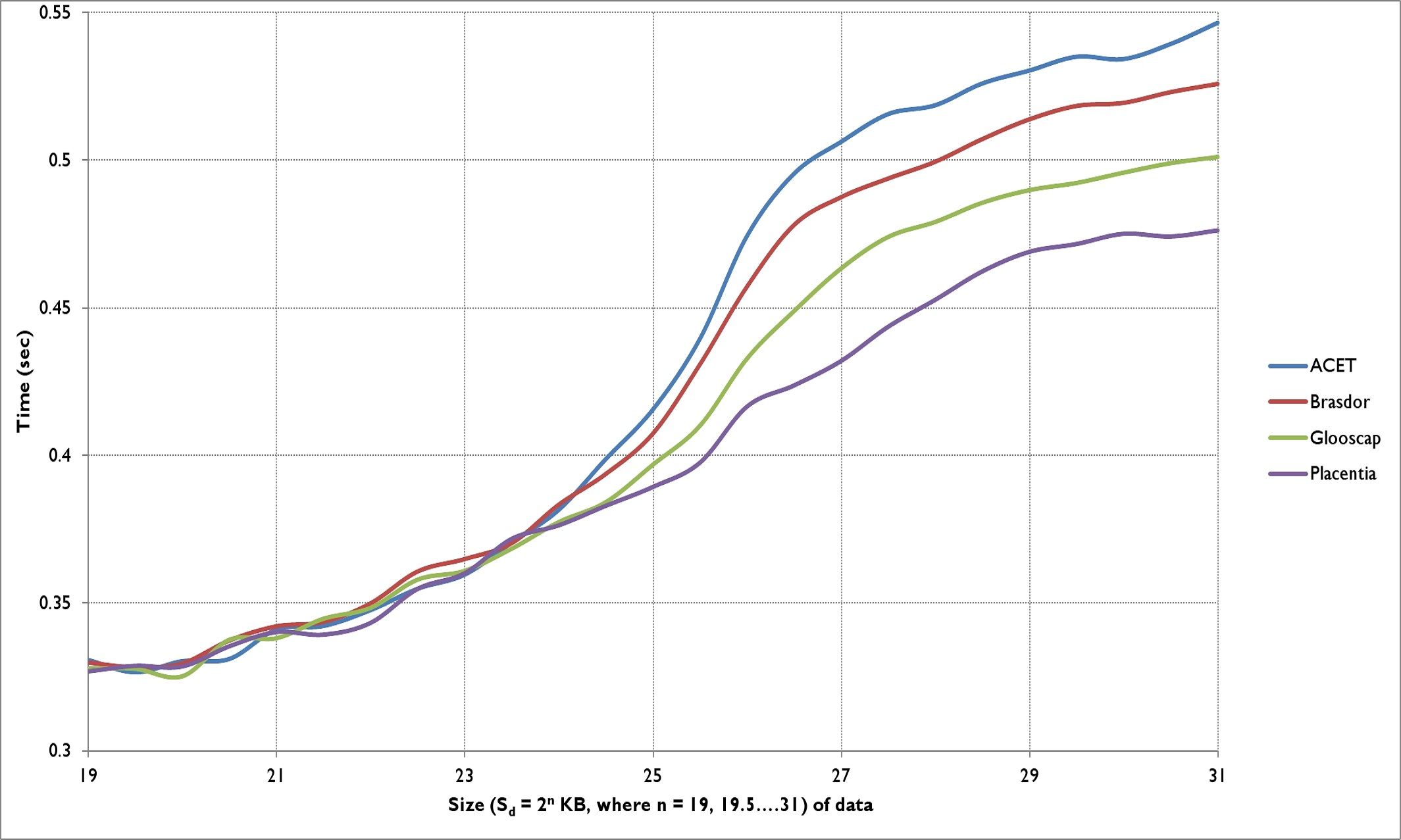}
	\caption{Size of data vs time taken for reinstating execution after failure in the agent intelligent approach}
	\label{figure10}
\end{figure}

\begin{figure}
	\centering
	\includegraphics[width=0.49\textwidth]{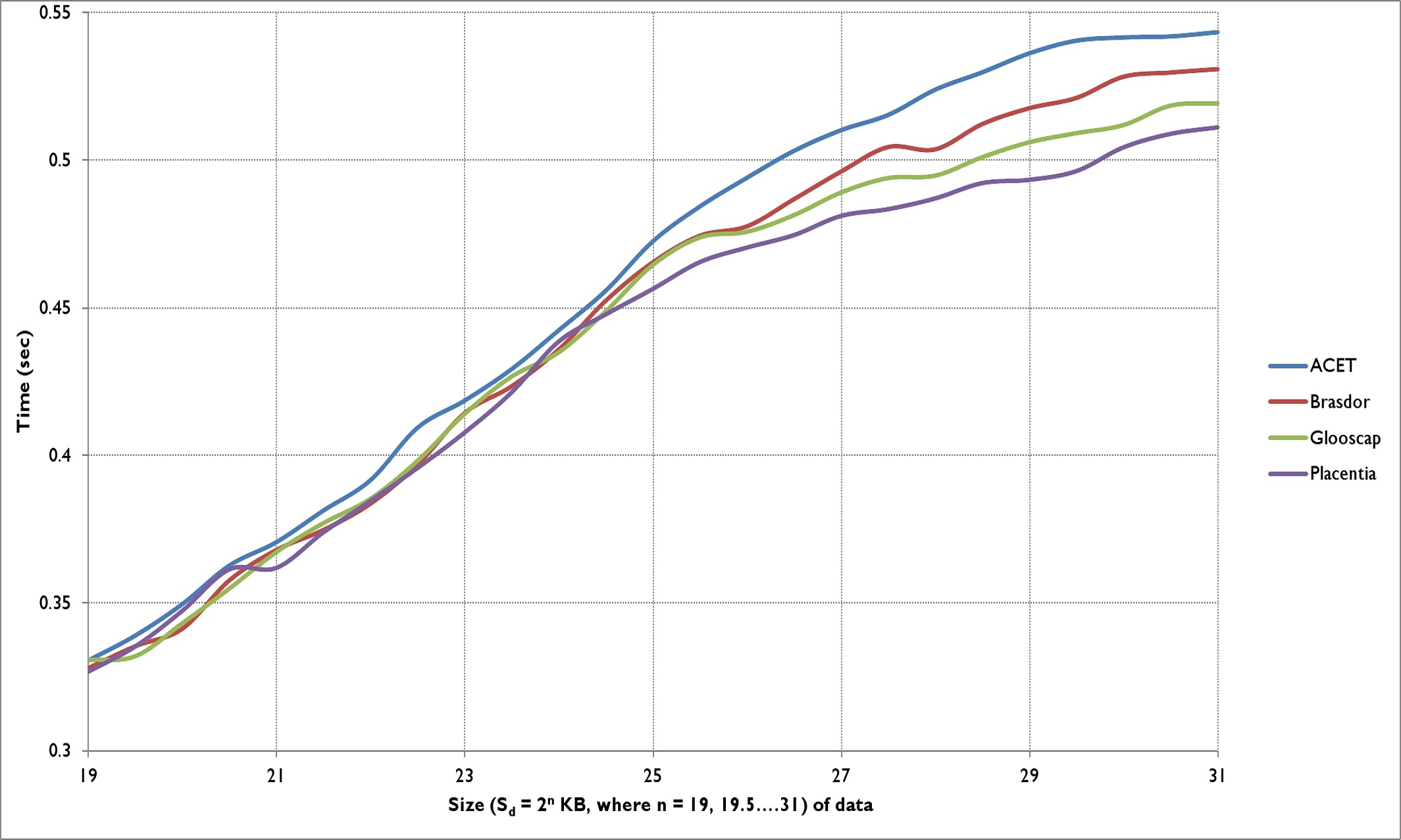}
	\caption{Size of data vs time taken for reinstating execution after failure in the core intelligent approach}
	\label{figure11}
\end{figure}

\begin{figure}
	\centering
	\includegraphics[width=0.49\textwidth]{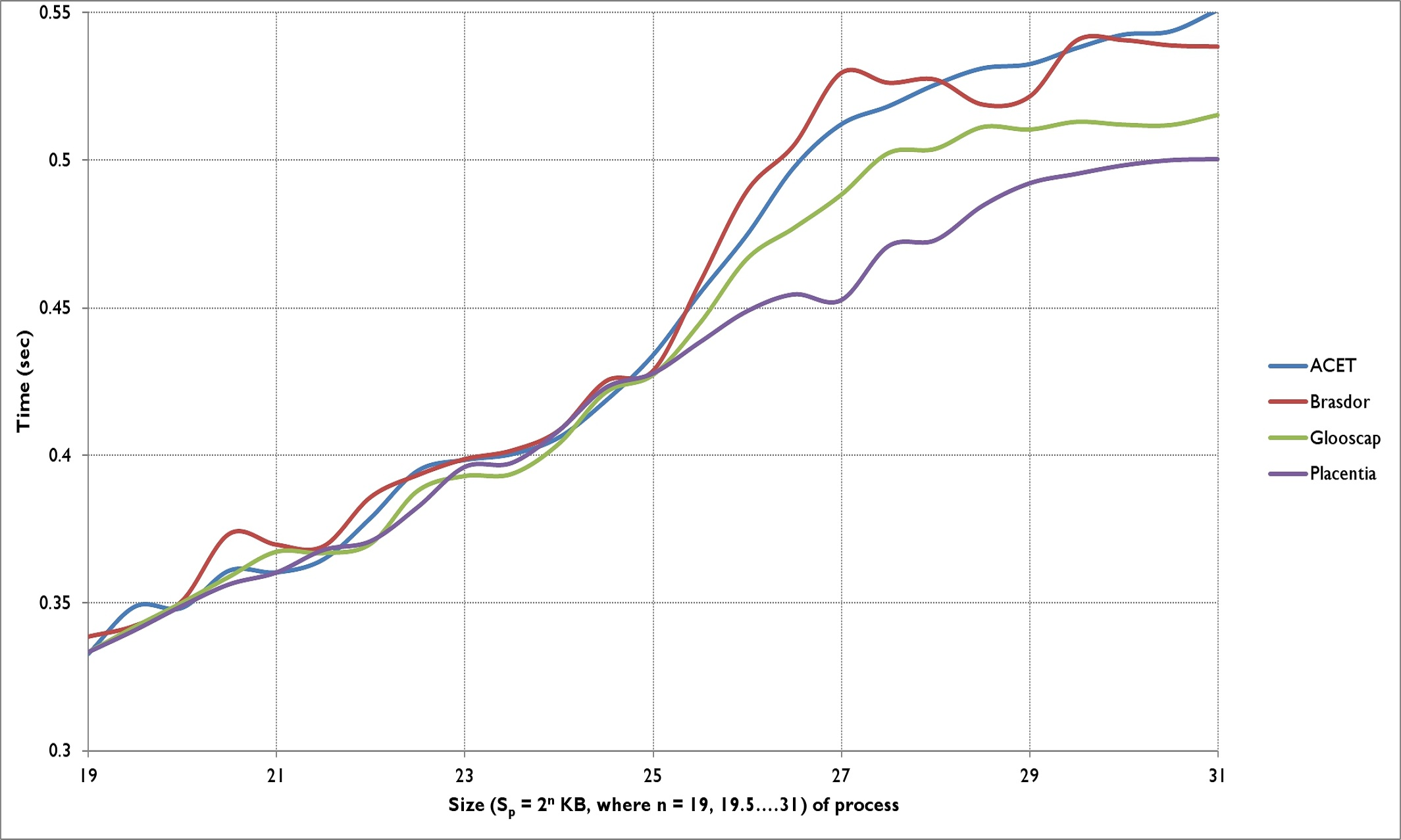}
	\caption{Process size vs time taken for reinstating execution after failure in the agent intelligent approach}
	\label{figure12}
\end{figure}

\begin{figure}
	\centering
	\includegraphics[width=0.49\textwidth]{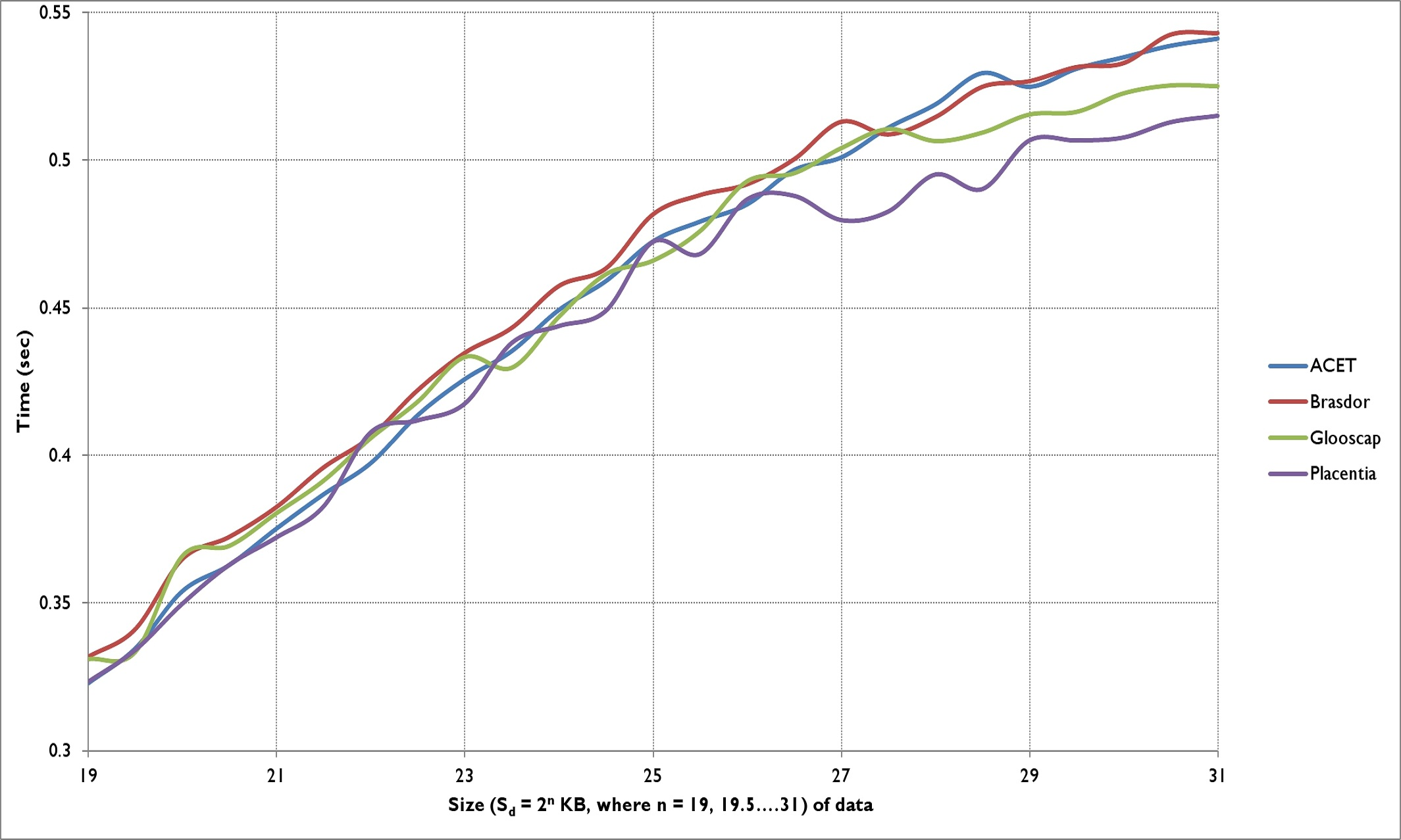}
	\caption{Process size vs time taken for reinstating execution after failure in the core intelligent approach}
	\label{figure13}
\end{figure}

\begin{itemize}
\item[(i)] The number of dependencies of the sub-job being executed denoted as $Z$. If the total number of input dependencies is $d_{i}$ and the total number of output dependencies is $d_{o}$, then $Z = d_{i} + d_{o}$. For example, in a parallel summation algorithm incorporating binary trees, each node has two input dependencies and one output dependency, and therefore $Z = 3$. In the experiments, the number of dependencies is varied between 3 and 63, by changing the number of input dependencies of an agent or a core. The results are presented in Figure 8 and Figure 9. 
\item[(ii)] The size of the data communicated across the cores denoted as $S_{d}$. In the experiments, the input data is a matrix for parallel summation and its size is varied between $2^{19}$ to $2^{31}$ KB. The results are presented in Figure 10 and Figure 11. 
\item[(iii)] The process size of the distributed components of the job denoted as $S_{p}$. In the experiments, the process size is varied between $2^{19}$ to $2^{31}$ KB which is proportional to the input data. The results are presented in Figure 12 and Figure 13. 
\end{itemize}

Figure 8 is a graph of the time taken in seconds for reinstating execution versus the number of dependencies in the agent intelligence failure scenario. The mean time taken to reinstate execution for 30 trials, ${{\Delta T}_{A}}_{2}$, is computed for varying numbers of dependencies, $Z$ ranging from 3 to 63. The size of the data on the agent is $S_{d}=2^{24}$ kilo bytes. The approach is slowest on the ACET cluster and fastest on the Placentia cluster. In all cases the communication overheads result in a steep rise in the time taken for execution until $Z=10$. The time taken on the ACET cluster rises once again after $Z=25$.

Figure 9 is a graph of the time taken in seconds for reinstating execution versus the number of dependencies in the core intelligence failure scenario. The mean time taken to reinstate execution for 30 trials, ${{\Delta T}_{C}}_{2}$, is computed for varying number of dependencies, $Z$ ranging from 3 to 63. The size of the data on the core is $S_{d}=2^{24}$ kilo bytes. The approach requires almost the same time on the four clusters for reinstating execution until $Z=10$, after which there is divergence in the plots. The approach lends itself well on Placentia and Glooscap.

Figure 10 is a graph showing the time taken in seconds for reinstating execution versus the size of data in kilobytes (KB), $S_{d} = 2^n$, where $n = 19, 19.5 \cdots 31$, carried by an agent in the agent intelligence failure scenario. The mean time taken to reinstate execution for 30 trials, ${{\Delta T}_{A}}_{2}$, is computed for varying sizes of data ranging from $2^{19}$ to $2^{31}$ KB. The number of dependencies $Z$ is 10 for the graph plotted. Placentia and Glooscap outperforms ACET and Brasdor in the agent approach for varying size of data.

Figure 11 is a graph showing the time taken in seconds for reinstating execution versus the size of data in kilobytes (KB), $S_{d} = 2^n$, where $n = 19, 19.5 \cdots 31$, on a core in the core intelligence failure scenario. The mean time taken to reinstate execution for 30 trials, ${{\Delta T}_{C}}_{2}$, is computed for varying sizes of data ranging from $2^{19}$ to $2^{31}$ KB. The number of dependencies $Z$ is 10 for the graph plotted. In this graph, nearly similar time is taken by the approach on the four clusters with the ACET cluster requiring more time than the other clusters for $n > 24$.

Figure 12 is a graph showing the time taken in seconds for reinstating execution versus process size in kilobytes (KB), $S_{p} = 2^n$, where $n = 19, 19.5 \cdots 31$, in the agent intelligence failure scenario. The mean time taken to reinstate execution for 30 trials, ${{\Delta T}_{A}}_{2}$, is computed for varying process sizes ranging from $2^{19}$ to $2^{31}$ KB. The number of dependencies $Z$ is 10 for the graph plotted. The second scenario performs similar to the first scenario. The approach takes almost similar times to reinstate execution after a failure on the four clusters, but there is a diverging behaviour after $n > 26$. 

Figure 13 is a graph showing the time taken in seconds for reinstating execution versus process size in kilobytes (KB), $S_{p} = 2^n$, where $n = 19, 19.5 \cdots 31$, in the core intelligence failure scenario. The mean time taken to reinstate execution for 30 trials, ${{\Delta T}_{C}}_{2}$, is computed for varying process sizes ranging from $2^{19}$ to $2^{31}$ KB. The number of dependencies $Z$ is 10 for the graph plotted. The approach has similar performance on the four clusters, though Placentia performs better than the other three clusters for a process size of more than $2^{26}$ KB.

\subsubsection{Decision Making Rules}

Parallel simulations in molecular dynamics model atoms or molecules in gaseous, liquid or solid states as point masses which are in motion. Such simulations are useful for studying the physical and chemical properties of the atoms or molecules. Typically the simulations are compute intensive and can be performed in at least three different ways \cite{26}. Firstly, by assigning a group of atoms to each processor, referred to as atom decomposition. The processor computes the forces related to the group of atoms to update their position and velocities. The communication between atoms is high and effects the performance on large number of processors. Secondly, by assigning a block of forces from the force matrix to be computed to each processor, referred to as force decomposition. This technique scales better than atom decomposition but is not a best solution for large simulations. Thirdly, by assigning a three dimensional space of the simulation to each processor, referred to as spatial decomposition. The processor needs to know the positions of atoms in the adjacent space to compute the forces of atoms in the space assigned to it. The interactions between the atoms are therefore local to the adjacent spaces. In the first and second decomposition techniques interactions are global and thereby dependencies are higher. 

Agent and core based approaches to fault tolerance can be incorporated within parallel simulations in the area of molecular dynamics. 
However, which of the two approaches, agent or core intelligence, is most appropriate? The decomposition techniques considered above establish dependencies between blocks of atoms and between atoms. Therefore the degree of dependency affects the relocation of a sub-job in the event of a core failure and reinstating it. The dependencies of an atom in the simulation can be based on the input received from neighbouring atoms and the output propagated to neighbouring atoms. Based on the number of atoms allocated to a core and the time step of the simulation the intensity of numerical computations and the data managed by a core vary. Large simulations that extend over long periods of time generate and need to manage large amounts of data; consequently the process size on a core will also be large. 

Therefore, (i) the dependency of the job, (ii) the data size and (iii) the process size are factors that need to be taken into consideration for deciding whether an agent-based approach or a core-based approach needs to come into play. Along with the observations from parallel simulations in molecular dynamics, the experimental results provide an insight into the rules for decision-making for large-scale applications. 

From the experimental results graphed in Figure 8 and Figure 9, where dependencies are varied, core intelligence is superior to agent intelligence if the total dependencies $Z$ is less than or equal to 10. Therefore, 
\begin{enumerate} [leftmargin = 1.5cm]
\item[Rule 1:]If the algorithm needs to incorporate fault tolerance based on the number of dependencies, then if $Z \leq 10$ use core intelligence, else use agent or core intelligence.
\end{enumerate}

From the experimental results graphed in Figure 10 and Figure 11, where the size of data is varied, agent intelligence is more beneficial than core intelligence if the size of data $S_{d}$ is less than or equal to $2^{24}$ KB. Therefore,
\begin{enumerate}[leftmargin = 1.5cm]
\item[Rule 2:]If the algorithm needs to incorporate fault tolerance based on the size of data, then if $S_{d} \leq 2^{24}$ KB, then use agent intelligence, else use agent or core intelligence.
\end{enumerate}

From the experimental results graphed in Figure 12 and Figure 13, where the size of the process is varied, agent intelligence is more beneficial than core intelligence if the size of the process $S_{p}$ is less than or equal to $2^{24}$ KB. Therefore,
\begin{enumerate}[leftmargin = 1.5cm]
\item[Rule 3:]If the algorithm needs to incorporate fault tolerance based on process size, then if $S_{p} \leq 2^{24}$ KB, then use agent intelligence, else use agent or core intelligence. 
\end{enumerate}

The number of dependencies, size of data, and process size are the three factors taken into account in the experimental results. The results indicate that the approach incorporating core intelligence takes lesser time than the approach incorporating agent intelligence. There are two reasons for this. Firstly, in the agent approach, the agent needs to establish the dependency with each agent individually, where as in the core approach as a job is migrated from a core onto another its dependencies are automatically established. Secondly, agent intelligence is a software abstraction of the sub-job, thereby adding a virtualised layer in the communication stack. This increases the time for communication. The virtual core is also an abstraction of the computing core but is closer to the computing core in the communication stack.

The above rules can be incorporated to exploit both agent-based and core-based intelligence in a third, hybrid approach. The key concept of the hybrid approach combines the mobility of the agents on the cores and the cores collectively executing a job. The approach can select whether the agent-based approach or the core-based approach needs to come to play based on the rules for decision-making. 

The key observation from the experimental results is that the cost of incorporating intelligence at the job and core levels for automating fault tolerance is less than a second, which is smaller than the time taken by manual methods which would be in the order of minutes. For example, in the first approach, the time for reinstating execution with over 50 dependencies is less than 0.55 seconds and in the second approach, is less than 0.5 seconds. Similar results are obtained when the size of data and the process are large.

\subsubsection{Genome Searching using Multi-Agent approaches}
The proposed multi-agent approaches and the decision making rules considered in the above sections are validated using a computational biology job. A job that fits the criteria of reduction algorithms is considered. In reduction algorithms, a job is decomposed to sub-jobs and executed on multiple nodes and the results are further passed onto other node for completing the job. One popular computational biology job that fits this criteria is searching for a genome pattern. This has been widely studied and fast and efficient algorithms have been developed for searching genome patterns (for example, \cite{gen01}, \cite{gen02} and \cite{gen03}). In the genome searching experiment performed in this research multiple nodes of a cluster execute the search operation and the output produced by the search nodes are then combined by an additional node.  

The focus of this experimental study is not parallel efficiency or scalability of the job but to validate the multi-agent approaches and the decision making rules in the context of computational biology. Hence, a number of assumptions are made for the genome searching job. First, redundant copies of the genome data are made on the same node to obtain a sizeable input. Secondly, the search operation is run multiple times to span long periods of time. Thirdly, the jobs are executed such that they can be stopped intentionally by the user at any time and gather the results of the preceding computations until the execution was stopped. 

The Placentia cluster is chosen for this validation study since it was the best performing cluster in the empirical study presented in the previous sections. The job is implemented using R programming which uses MPI for exploiting computation on multiple nodes of the Placentia cluster. Bioconductor packages\footnote{http://bioconductor.org/} are required for supporting the job. The job makes use of BSgenome.Celegans.UCSC.ce2, BSgenome.Celegans.UCSC.ce6 and BSgenome.Celegans.UCSC.ce10 as input data which are the ce2, ce6 and ce10 genome for chromosome I of Caenorhabditis elegans \cite{gen04, gen05}. A list of 5000 genome patterns each of which is a short nucleotide sequence of 15 to 25 bases is provided to be searched against the input data. 

The forward and reverse strands of seven Caenorhabditis elegans chromosomes named as chrI, chrII, chrIII, chrIV, chrV, chrX, chrM are the targets of the search operation. When there is a target hit the search nodes provide to the node that gathers the results the name of the chromosome where the hit occurs, two integers giving the starting and ending positions of the hit, an indication of the hit either in the forward or reverse strand, and unique identification for every pattern in the dictionary. The results are tabulated in an output file in the combining node. A sample of the output is shown in Figure 14. 

\begin{figure}
	\centering
	\includegraphics[width=0.45\textwidth]{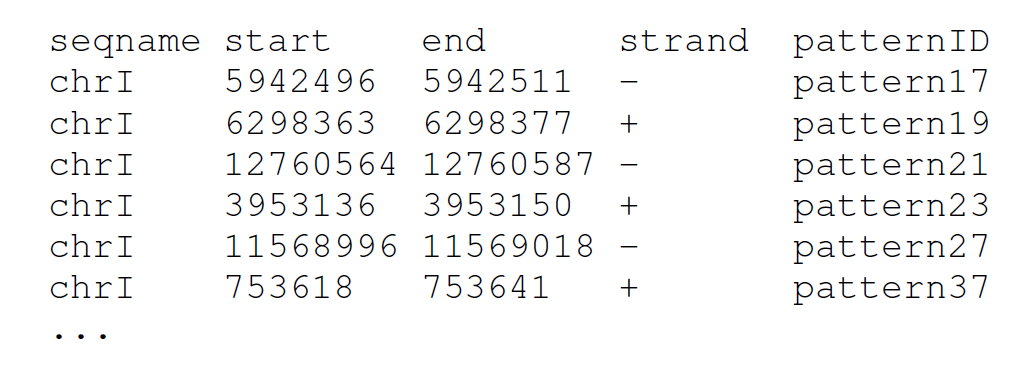}
	\caption{Sample output from searching genome pattern. The output shows the name of the chromosome where the target hit occurs, followed by two integers giving the starting and ending positions of the hit, an indication of the hit either in the forward or reverse strand, and unique identification for every pattern in the dictionary.}
	\label{figure14}
\end{figure}

Redundant copies of the input data are made to obtain 512 MB (which is $2^{19}$ KB) and the job is executed for one hour. In a typical experiment the number of dependencies, $Z$ was set to 4; three nodes of the cluster performed the search operation while the fourth node combined the results passed on to it from the three search nodes. In the agent intelligence based approach the time for predicting the fault is 38 seconds, the time for reinstating execution is 0.47 seconds, the overhead time is over 5 minutes and the total time when one failure occurs per hour is 1 hour, 6 minutes and 17 seconds. In the core intelligence based approach the time for predicting the single node failure is similar to the agent intelligence approach; the time for reinstating execution is 0.38 seconds, the overhead time is over 4 minutes and the total time when one failure occurs per hour is 1 hour, 5 minutes and 8 seconds. 

In another experiment for 512 MB size of input data the number of dependencies was set to 12; eleven nodes for searching and one node for combining the results provided by the eleven search nodes. In the agent intelligence based approach the time for reinstating execution is 0.54 seconds, the overhead time is over 6 minutes and the total time when one failure occurs per hour is 1 hour, 7 minutes and 34 seconds. In the core intelligence based approach the time for reinstating execution is close to 0.54 seconds, the overhead time is over 6 minutes and the total time when one failure occurs per hour is 1 hour, 7 minutes and 48 seconds.  

The core intelligence approach requires less time than the agent intelligence approach when $Z=3$, but the times are comparable when $Z=12$. So, the above two experiments validate Rule 1 for decision making considered in the previous section.

Experiments were performed for different input data sizes; in one case $S_{d} = 2^{19}$ KB and in the other $S_{d} = 2^{25}$ KB. The agent intelligence approach required less time in the former case than the core intelligence approach. The time was comparable for the latter case. Hence, the genome searching job in the context of the experiments validated Rule 2 for decision making. Similarly, when process size was varied Rule 3 was found to be validated.     

The genome searching job is used as an example to validate the use of the multi-agent approaches for computational biology jobs. The decision making rules empirically obtained were satisfied in the case of this job. The results obtained from the experiments for the genome searching job along with comparisons against traditional fault tolerance approaches, namely centralised and decentralised checkpointing are considered in the next section.

\section{Discussion}
\label{discussion}

All fault tolerance approaches initiate a response to address a failure. Based on when a response is initiated with respect to the occurrence of the failure, approaches can be classified as proactive and reactive. Proactive approaches predict failures of computing resources before they occur and then relocate a job executing on resources anticipated to fail onto resource that are not predicted to fail (for example \cite{64, 70, 71}). Reactive approaches on the other hand minimise the impact of failures after they have occurred (for example checkpointing \cite{09}, rollback recovery \cite{72} and message logging \cite{73}). A hybrid of proactive and reactive, referred to as adaptive approaches, is implemented so that failures that cannot be predicted by proactive approaches are handled by the reactive approaches \cite{74, 75, 76}.

The control of a fault tolerant approach can be either centralised or distributed. In approaches where the control is centralised, one or more servers are used for backup and a single process responsible for monitoring jobs that are executed on a network of nodes. The traditional message logging and checkpointing approach involves the periodic recording of intermediate states of execution of a job to which execution can be returned if faults occur. Such approaches are susceptible to single point failure, lack scalability over a large network of nodes, have large overheads, and require large disk storage. These drawbacks can be minimised or avoided when the control of the approaches is distributed (for example, distributed diagnosis \cite{77}, distributed checkpointing \cite{68} and diskless checkpointing \cite{78}). 

In this paper two distributed proactive approaches towards achieving fault tolerance are proposed and implemented. In both approaches a job to be computed is decomposed into sub-jobs which are then mapped onto the computing cores. The two approaches operate at the middle levels (between the sub-jobs and the computing cores) incorporating agent intelligence. In the first approach, the sub-jobs are mapped onto agents which are released onto the cores. If an agent is notified of a potential core failure during execution of the sub-job mapped onto it, then the agent moves onto another core thereby automating fault tolerance. In the second approach the sub-jobs are scheduled on virtual cores, which are an abstraction of the computing cores. If a virtual core anticipates a core failure then it moves the sub-job on it to another virtual core, in effect onto another computing core. The two approaches achieve automation in fault tolerance using intelligence in agents and using intelligence in cores respectively. A third approach is proposed which brings together the concepts of both agent intelligence and core intelligence from the first two approaches. 

\subsection{Overcoming the problems of Checkpointing}
The conventional approaches to fault tolerance such as checkpointing have large communication overheads based on the periodicity of checkpointing. High frequencies of checkpointing can lead to heavy network traffic since the available communication bandwidth will be saturated with data transferred from all computing nodes to the a stable storage system that maintains the checkpoint. This traffic is on top of the actual data flow of the job being executed on the network of cores. While global approaches are useful for jobs which are less memory and data intensive and can be executed over short periods of time, they may constrain the efficiency for jobs using big data in limited bandwidth platforms. Hence, local approaches can prove useful. In the case of the agent based approaches there is high periodicity for probing the cores in the background but very little data is transferred while probing unlike in checkpointing. Hence, communication overhead times will be significantly lesser. 

Lack of scalability is another issue that affects efficient fault tolerance. Many checkpointing strategies are centralised (with few exceptions, such as \cite{68, 69}) thereby limiting the scale of adopting the strategy. This can be mitigated by using multiple centralised checkpointing servers but the distance between the nodes and the server discounts the scalability issue. In the agent based approaches, all communications are short distance since the cores only need to communicate with the adjacent cores. Local communication therefore increases the scale to which the agent based approaches can be applied.

Checkpointing is susceptible to single point failures due to the failure of the checkpoint servers. The job executed will have to be restarted. The agent-based approaches are also susceptible to single point failures. While they incorporate intelligence to anticipate hardware failure the processor core may fail before the sub-job it supports can be relocated to an adjacent processor core, before the transfer is complete, or indeed the core onto which it is being transferred may also fail. However, the incorporation of intelligence on the processor core, specifically the ability to anticipate hardware failure locally, means that the numbers of these hardware failures that lead to job failure can be reduced when compared to traditional checkpointing. But since there is the possibility of agent failure the retention of some level of human intervention is still required. Therefore, we propose combining checkpointing with the agent-based approaches, the latter acting as a first line of anticipatory response to hardware failure backed up by traditional checkpointing as a second line of reactive response.


\subsection{Predicting potential failures}
Figure 15 shows the execution of a job between two checkpoints, $C_{n}$ and $C_{n+1}$, where $PF$ is the predicted failure and $F$ is the actual failure of the node on which a sub-job is executing. Figure 15(a) shows when there are no predicted failures or actual failures that occur on the node. Figure 15(b) shows when a failure occurs but could not be predicted. In this case, the system fails if the multi-agent approaches are solely employed. One way to mitigate this problem is by employing the multi-agent approaches in conjunction with checkpointing as shown in the next section. Figure 15(c) shows when the approaches predict a failure which does not happen. If a large number of such predictions occur then the sub-job needs to be shifted often from one node to the other which adds to the overhead time for executing the job. This is not an ideal case and makes the job unstable. Figure 15(d) shows the ideal case in which a fault is predicted before it occurs. 

\begin{figure}
	\centering
	\includegraphics[width=0.4\textwidth]{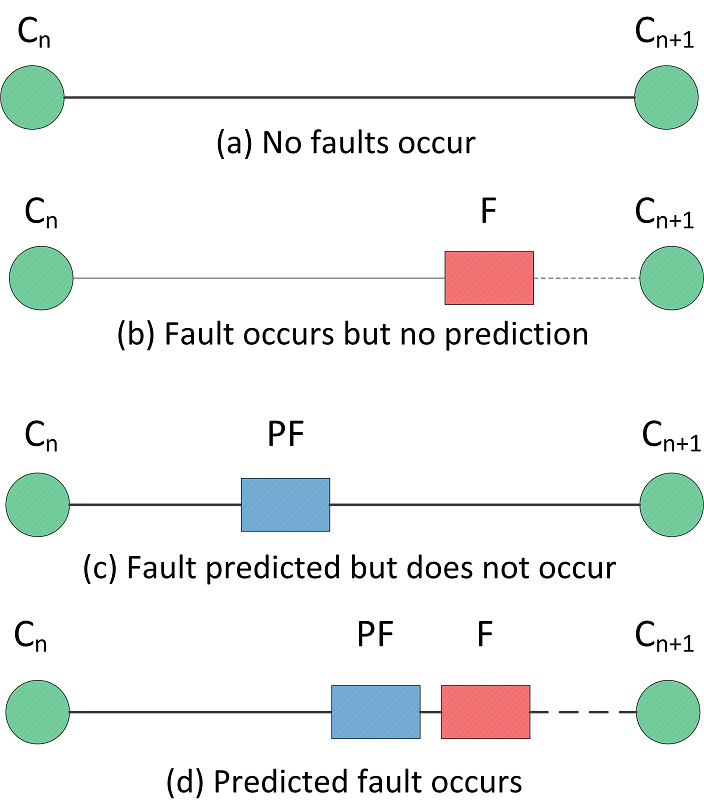}
	\caption{Fault prediction between two checkpoints, $C_{n}$ and $C_{n+1}$. (a) Ideal state of the job when no faults occur. (b) Failure state of the job when a fault occurs but is not predicted. (c) Unstable state of the job when a fault is predicted but does not occur. (d) Ideal prediction state of the job when a fault is predicted and occurs thereafter.}
	\label{figure15}
\end{figure}

Failure prediction is based on a machine learning approach that is incorporated within multi-agents. This prediction is based on a log that is maintained on the health of the node and its adjacent nodes. Each agent sends out 'are you alive' signals to adjacent nodes to gather the state of the adjacent node. The machine learning approach is constantly evaluating the state of the system against the log it maintains, which is different across the nodes. The log can contain the state of the node from past failures, work load of the nodes when it failed previously and even data related to patterns of periodic failures. However, this prediction method cannot predict a range of faults due to deadlocks, hardware and power failures and instantaneously occurring faults. Hence, the multi-agent approaches are most useful when used along with checkpointing. 

It was observed that nearly 29\% of all faults occurring in the cluster could be predicted. Although this number is seemingly small it is helpful to not have to rollback to a previous checkpoint when a large job under time constraints is executed. Accuracy of the predictions were found to be 64\%; the system was found to be stable in 64 out of the 100 times a prediction was made. To increase the impact of the multi-agent approaches more faults will need to be captured. For this extensive logging and faster methods for prediction will need to be considered. These approaches will have to be used in conjunction with checkpointing for maximum effectiveness. The instability due to the approaches shifting jobs between nodes when there is a false prediction will need to be reduced to improve the overall efficiency of the approaches. For this, the state of the node can be compared with other nodes so that a more informed choice is made by the approaches.  

\subsection{Comparing traditional and multi-agent approaches}
Table 1 shows a comparison between a number of fault tolerant strategies, namely centralised and decentralised checkpointing and the multi-agent approaches. An experiment was run for a genome searching job that was executed multiple times on the Placentia cluster. Data in the table was obtained to study the execution of the genome searching job between two checkpoints ($C_{n}$ and $C_{n+1}$) which are one hour apart. The execution is interrupted by failure $F$ as shown in Figure 16. Two types of single node failure are simulated in the execution. The first is a periodic node failure which occurs at 15 minutes after $C_{n}$ and 45 minutes before $C_{n+1}$ (refer Figure 16(a)), and the second is a random node failure which occurs $x$ minutes after $C_{n}$ and $60-x$ minutes before $C_{n+1}$ (refer Figure 16(b)). The average time when a random failure occurs is found to be 31 minutes and 14 seconds for 5000 trials. The size of data, $S_{d} = 2^{19}$ KB and the number of dependencies, $Z=4$. 

\begin{figure*}
	\centering
	\includegraphics[width=0.9\textwidth]{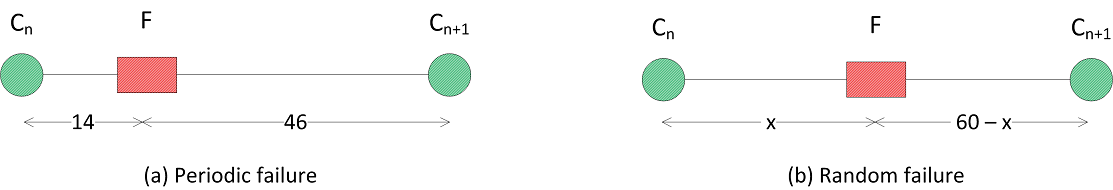}
	\caption{Fault occurrences between two checkpoints, $C_{n}$ and $C_{n+1}$. (a) Periodic failure that occurs 14 minutes after $C_{n}$ and 46 minutes before $C_{n+1}$. (b) Random failure that occurs $x$ minutes after $C_{n}$ and $60-x$ minutes before $C_{n+1}$.}
	\label{figure16}
\end{figure*}

In Table 1, the average time taken for reinstating execution, for the overheads and for executing the job between the checkpoints is considered. The time taken for reinstating execution is for bringing execution back to normal after a failure has occurred. The reinstating time is obtained for one periodic single node failure and one random single node failure. The overhead time is for creating the checkpoints and transferring data for the checkpoint to the server. The overhead time is obtained for one periodic single node failure and one random single node failure. The execution time without failures, when one periodic failure occurs per hour and when five random failures occur per hour is obtained.

Centralised checkpointing using single and multiple servers is considered when the frequency of checkpointing is once every hour. In the case of both single and multiple server checkpointing the time taken for reinstating execution regardless of whether it was a periodic or random failure is 14 minutes and 8 seconds. On a single server the overhead is 8 minutes and 5 seconds where as the overhead to create the checkpoint is 9 minutes and 14 seconds which is higher than overheads on a single server and is expected. The average time taken for executing the job when one failure occurs includes the elapsed execution time (15 minutes for periodic failure and 31 minutes and 14 seconds for random failure) until the failure occurred and the combination of the time for reinstating execution after the failures and the overhead time. For one periodic failure that occurs in one hour the penalty of execution when single server checkpointing is 62\% more than executing without a failure; in the case of a random failure that occurs in one hour the penalty is 89\% more than executing without a failure. If five random failure occur then the penalty is 445\%, requiring more than five times the time for executing the job without failures. 

Centralised checkpointing with multiple servers requires more time than with a single server. This is due to the increase in the overhead time for creating checkpoints on multiple servers. Hence, checkpointing with multiple servers requires 64\% and 91\% more time than the time for executing the job without any failures for one periodic and one random failure per hour respectively. On the other hand executing jobs when decentralised checkpointing on multiple servers is employed requires similar time to that taken by centralised checkpointing on a single server. The time for reinstating execution is higher than centralised checkpointing methods due to the time required for determining the server closest to the node that failed. However, the overhead times are lower than other checkpoint approaches since the server closest to the node that failed is chosen for creating the checkpoint which reduces data transfer times.  

The multi-agent approaches are proactive and therefore the average time taken for predicting single node failures are taken into account which is nearly 38 seconds. The time taken for reinstating execution after one periodic single node failure for the agent intelligence approach is 0.47 seconds and for the core intelligence approach is 0.38 seconds. Since $Z \leq 10$ the core intelligence approach is selected. In this case, the core intelligence approach is faster than the agent intelligence approach in the total time it takes for executing the job when there is one periodic or random fault and when there are five faults that occur in the job. The multi-agent approaches only require one-fifth the time taken by the checkpointing methods for completing execution. This is because the time for reinstating and the overhead times are significantly lower than the checkpointing approaches.  

Table 2 shows a comparison between centralised and decentralised checkpointing and the multi-agent approaches for a genome searching job that is executed on the Placentia cluster for five hours. The checkpoint periodicity is once every one, two and four hours as shown in Figure 17. Similar to Table 1 periodic and random failures are simulated. Figure 17(a) shows the start and completion of the job without failures or checkpoints. When the checkpoint periodicity is one hour there are four checkpoints, $C_{1}$, $C_{2}$, $C_{3}$ and $C_{4}$ (refer Figure 17(b)); a periodic node failure is simulated after 14 minutes from a checkpoint and the average time at which a random node failure occurs is found to be 31 minutes and 14 seconds from a checkpoint for 5000 trials. When checkpoint periodicity is two hours there are two checkpoints, $C_{1}$ and $C_{2}$ (refer Figure 17(c)); a periodic node failure is simulated after 28 minutes from a checkpoint and the average time a random node failure occurs is found to be after 1 hour, 3 minutes and 22 seconds from a checkpoint for 5000 trials. When checkpoint periodicity is four hours there is only one checkpoint $C_{1}$ (refer Figure 17(d)); a periodic node failure is simulated after 56 minutes from a checkpoint and the average time at which a random failure occurs is found to be after 2 hours, 8 minutes and 47 seconds from each checkpoint for 5000 trials. 

\begin{figure*}
	\centering
	\includegraphics[width=0.75\textwidth]{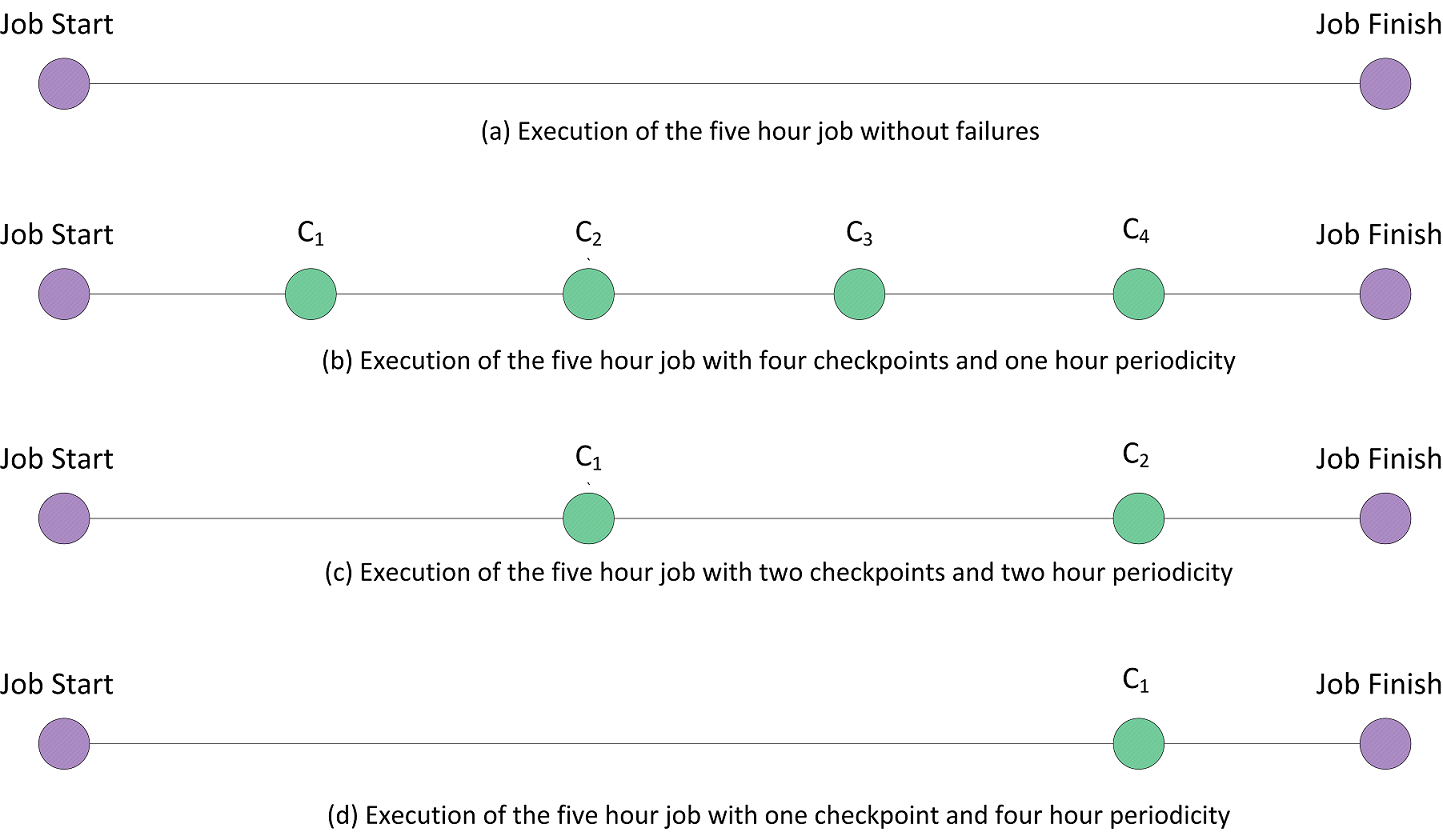}
	\caption{Execution of the five hour job with and without checkpoints. (a) When no checkpoints are placed from the start to completion of the job. (b) When four checkpoints each one hour apart are placed from start to completion of the job. (c) When two checkpoints each two hours apart are placed from start to completion of job. (d) When one checkpoint is placed after four hours of starting the job.}
	\label{figure17}
\end{figure*}

Similar to Table 1, in Table 2, the average time taken for reinstating execution, for the overheads and for executing the job from the start to finish with and without checkpoints is considered. The time to bring execution back to normal after a failure has occurred is referred to as reinstating time. The time to create checkpoints and transfer checkpoint data to the server is referred to as the overhead time. The execution of the job when one periodic and one random failure occurs per hour and when five random failures occur per hour is considered. 

Without checkpointing the genome searching job is run such that a human administrator monitors the job from its start until completion. In this case, if a node fails then the only option is to restart the execution of the job. Each time the job fails and given that the administrator detected it using cluster monitoring tools as soon as the node failed approximately, then at least ten minutes are required for reinstating the execution. If a periodic failure occurred once every hour from the 14th minute from execution then there are five periodic faults. Once a failure occurs the execution will always have to come back to its previous state by restarting the job. Hence, the five hour job, with just one periodic failure occurring every hour will take over 21 hours. Similarly, if a random failure occurred once every hour (average time of occurrence is 31 minutes and 14 seconds after execution starts), then there are five failure points, and over 23 hours are required for completing the job. When five random failures occur each hour of the execution then more than 80 hours are required; this is nearly 16 times the time for executing the job without a failure.  

Centralised checkpointing on a single server and on multiple servers and decentralised checkpointing on multiple servers for one, two and four hour periodicity in the network are then considered in Table 2. For checkpointing methods when one hour frequency is chosen more than five times the time taken for executing the job without failures is required. When the frequency of checkpointing is every two hours then just under four times the time taken for executing the job without failures is required. In the case when the checkpoint is created every four hours just over 3 times the time taken for executing the job without failures is required. The multi-agent approaches on the other hand take only one-fourth the time taken by traditional approaches for the job with five single node faults that occur each hour. This is significant time saving for running jobs that require many hours for completing execution.


\subsection{Similarities and differences between the approaches}
The agent and core intelligence approaches are similar in at least four ways. Firstly, the objective of both the approaches is to automate fault tolerance. Secondly, the job to be executed is broken down into sub-jobs which are executed. Thirdly, fault tolerance is achieved in both approaches by predicting faults likely to occur in the computing core. Fourthly, technology enabling mobility is required by both the approaches to carry the sub-job or to push the sub-job from one core onto another. These important similarities enable both the agent and core approaches to be brought together to offer the advantages as a hybrid approach. 

While there are similarities between the agent and core intelligence approaches there are differences that reflect in their implementation. These differences are based on: (i) Where the job is situated - in the agent intelligence approach, the sub-job becomes the payload of an agent situated on a computing core. In the core intelligence approach, the sub-job is situated on a virtual core, which is an abstraction of the computing core. (ii) Who predicts the failures - the agent constantly probes the compute core it is situated on and predicts failure in the agent approach, whereas in the core approach the virtual core anticipates the failure. (iii) Who reacts to the prediction - the agent moves onto another core and re-establishes its dependencies in the agent approach, whereas the virtual core is responsible for moving a sub-job onto another core in the core approach. (iv) How dependencies are updated - an agent requires to carry information of its dependencies when it moves off onto another core and establishes its dependencies manually in the agent approach, whereas the dependencies of the sub-job on the core do not require to be manually updated in the core approach. (v) What view is obtained - in the agent approach, agents have a global view as they can traverse across the network of virtual cores, which is in contrast to the local view of the virtual cores in the core approach.

\section{Conclusions}
\label{conclusions}

The agent based approaches described in this paper offer a candidate solution for automated fault tolerance or in combination with checkpointing as proposed above offer a means of reducing current levels of human intervention. The foundational concepts of the agent and core based approaches were validated on four computer clusters using parallel reduction algorithms as a test case in this paper. Failure scenarios were considered in the experimental studies for the two approaches. The effect of the number of dependencies of a sub-job being executed, the volume of data communicated across cores, and the process size are three factors considered in the experimental studies for determining the performance of the approaches. 

The approaches were studied in the context of parallel genome searching, a popular computational biology job, that fits the criteria of a parallel reduction algorithm. The experiments were performed for both periodic and random failures. The approaches were compared against centralised and decentralised checkpointing approaches. In a typical experiment in which the fault tolerant approaches are studied in between two checkpoints one hour apart when one random failure occurs, centralised and decentralised checkpointing on an average add 90\% to the actual time for executing the job without any failures. On the other hand, in the same experiment the multi-agent approaches add only 10\% to the overall execution time. The multi-agent approaches cannot predict all failures that occur in the computing nodes. Hence, the most efficient way of incorporating these approaches is to use them on top of checkpointing. The experiments demonstrate the feasibility of such approaches for computational biology jobs. The key result is that a job continues execution after a core has failed and the time required for reinstating execution is lesser than checkpointing methods. 

Future work will explore methods to improve the accuracy of prediction as well as increase the number of faults that can be predicted using the multi-agent approaches. The challenge to achieve this will be to mine log files for predicting a wide range of faults and predict them as quickly as possible before the fault occurs. Although the approaches can reduce human administrator intervention they can be used independently only if a wider range of faults can be predicted with greater accuracy. Until then the multi-agent approaches can be used in conjunction with checkpointing for improving fault tolerance.


\begin{acknowledgments}
The authors would like to thank the administrators of the compute resources at the Centre for Advanced Computing and Emerging Technologies (ACET), University of Reading, UK and the Atlantic Computational Excellence Network (ACEnet). 
\end{acknowledgments}







\end{article}

\begin{sidewaystable}
\caption{\textbf{Comparing fault tolerant approaches between checkpoints with one hour periodicity.} The average time taken (a) for predicting one single node failure, (b) for reinstating execution after one periodic single node failure, (c) for reinstating execution after one random single node failure, (d) for the overheads related to one periodic single node failure, (e) for the overheads related to one random single node failure, and (f) for executing the job without failures and checkpoints, with one periodic failure per hour and with five periodic failures per hour are tabulated for centralised checkpointing with single and multiple servers, decentralised checkpointing with multiple servers and the multi-agent approaches.}
\label{table1}
\begin{tabular}{| p{3.8cm} | p{1.6cm} | p{1.8cm} | p{1.8cm} | p{1.6cm} | p{1.6cm} | p{1.5cm} | p{1.5cm} | p{1.5cm} | p{1.5cm} |}
\hline
\multirow{3}{*}{Fault tolerant approach}	&	\multicolumn{9}{c |}{Average time (hh:mm:ss) for}\\
\cline{2-10}
&	Predicting one single node failure	&	Reinstating execution after one periodic single node failure	&	Reinstating execution after one random single node failure	& 	Overheads related to one periodic single node failure	& 	Overheads related to one random single node failure	&	\multicolumn{4}{c |}{Executing job}\\
\cline{7-10}
&	&	&	&	&	&	Without failures and checkpoints	&	With one periodic failure per hour	&	With one random failure per hour	& With five random failures per hour\\
\hline
\hline
\multicolumn{10}{| l |}{Centralised checkpointing with single server}\\
\hline
1 hour periodicity	&	-	&	00:14:08	&	00:14:08	&	00:08:05	&	00:08:05	&	01:00:00	&	01:37:13	&	01:53:27	&	05:27:15\\
\hline
\multicolumn{10}{| l |}{Centralised checkpointing with multiple servers}\\
\hline
1 hour periodicity	& 	-	&	00:14:08	&	00:14:08	&	00:09:14	&	00:09:14	&	01:00:00	&	01:38:22	&	01:54:36	&	05:33:00\\
\hline
\multicolumn{10}{| l |}{Decentralised checkpointing on multiple servers}\\
\hline
1 hour periodicity	& 	-	&	00:15:27	&	00:15:27	&	00:06:44	&	00:06:44	&	01:00:00	&	01:37:11	&	01:53:25	&	05:27:05\\
\hline
\multicolumn{10}{| l |}{Multi-agent approaches}\\
\hline
Agent intelligence	& 	00:00:38	&	00:00:0.47	&	00:00:0.47	&	00:05:14	&	00:05:14	&	\multirow{3}{*}{01:00:00}	&	01:06:17	&	01:06:17	&	01:32:27\\
Core intelligence	& 	00:00:38	&	00:00:0.38	&	00:00:0.38	&	00:04:27	&	00:04:27	&	&	01:05:08	&	01:05:08	&	01:25:42\\
Hybrid intelligence	& 	00:00:38	&	00:00:0.38	&	00:00:0.38	&	00:04:27	&	00:04:27	&	&	01:05:08	&	01:05:08	&	01:25:42\\
\hline
\end{tabular}

\end{sidewaystable}

\begin{sidewaystable}
\caption{\textbf{Comparing fault tolerant approaches for a five hour job with checkpoints having one, two and four hour periodicity.} The average time taken (a) for predicting one single node failure, (b) for reinstating execution after one periodic single node failure, (c) for reinstating execution after one random single node failure, (d) for all the overheads related to one periodic single node failure, (e) for all the overheads related to one random single node failure, and (f) for executing the job without failures and checkpoints, with one periodic failure per hour and with five periodic failures per hour are tabulated for cold restart with no fault tolerance, centralised checkpointing with single and multiple servers, decentralised checkpointing with multiple servers and the multi-agent approaches.}
\label{table2}
\begin{tabular}{| p{3.8cm} | p{1.8cm} | p{1.8cm} | p{1.8cm} | p{1.8cm} | p{1.8cm} | p{1.2cm} | p{1.2cm} | p{1.2cm} | p{1.2cm} |}
\hline
\multirow{3}{*}{Fault tolerant approach}	&	\multicolumn{9}{c |}{Average time (hh:mm:ss) for}\\
\cline{2-10}
&	Predicting one single node failure	&	Reinstating execution after one periodic single node failure	&	Reinstating execution after one random single node failure	& 	All overheads related to one periodic single node failure	& 	All overheads related to one random single node failure	&	\multicolumn{4}{c |}{Executing job}\\
\cline{7-10}
&	&	&	&	&	&	Without failures	&	With one periodic failure per hour	&	With one random failure per hour	&	With five random failures per hour\\
\hline
\hline

Cold restart with no failure tolerance	&	-	&	00:10:00	&	00:10:00	&	-	&	-	&	05:00:00	&	21:15:17	&	23:01:00	&	80:31:04\\
\hline
\multicolumn{10}{| l |}{Centralised checkpointing with single server}\\
\hline
1 hour periodicity	& 	-	&	00:14:08	&	00:14:08	&	00:08:05	&	00:08:05	&	\multirow{3}{*}{05:00:00}	&	08:01:05	&	09:27:15	&	27:16:15\\
2 hour periodicity	& 	-	&	00:15:40	&	00:15:40	&	00:10:17	&	00:10:17	&				&	07:41:51	&	07:58:38	&	19:53:10\\
4 hour periodicity	& 	-	&	00:16:27	&	00:16:27	&	00:11:53	&	00:11:53	&				&	06:24:20	&	07:37:07	&	18:05:35\\
\hline
\multicolumn{10}{| l |}{Centralised checkpointing with multiple servers}\\
\hline
1 hour periodicity	& 	-	&	00:14:08	&	00:14:08	&	00:09:14	&	00:09:14	&	\multirow{3}{*}{05:00:00}	&	08:07:14	&	09:33:23	&	27:45:00\\
2 hour periodicity	& 	-	&	00:15:40	&	00:15:40	&	00:12:22	&	00:12:22	&				&	07:47:52	&	08:07:18	&	20:01:16\\
4 hour periodicity	& 	-	&	00:16:27	&	00:16:27	&	00:13:57	&	00:13:57	&				&	07:04:28	&	07:52:27	&	18:45:22\\
\hline
\multicolumn{10}{| l |}{Decentralised checkpointing on multiple servers}\\
\hline
1 hour periodicity	& 	-	&	00:15:27	&	00:15:27	&	00:06:44	&	00:06:44	&	\multirow{3}{*}{05:00:00}	&	08:00:55	&	09:27:05	&	27:15:25\\
2 hour periodicity	& 	-	&	00:17:23	&	00:17:23	&	00:09:46	&	00:09:46	&				&	07:40:18	&	07:57:36	&	19:48:00\\
4 hour periodicity	& 	-	&	00:18:33	&	00:18:33	&	00:13:03	&	00:13:03	&				&	06:27:36	&	07:40:23	&	18:21:55\\
\hline
\multicolumn{10}{| l |}{Agent intelligent approach}\\
\hline
1 hour periodicity	&	\multirow{3}{*}{00:00:38}	&	\multirow{3}{*}{00:00:0.47}	&	\multirow{3}{*}{00:00:0.47}	&	00:05:14	&	00:05:14	&	\multirow{3}{*}{05:00:00}	&	05:31:14	&	05:31:14	&	07:37:44\\
2 hour periodicity	&	&		&		&	00:06:38	&	00:06:38	&								&	05:20:34	&	05:20:34	&	06:42:41\\	
4 hour periodicity	&	&		&		&	00:07:41	&	00:07:41	&							&	05:16:27	&	05:16:27	&	05:39:16\\	
\hline
\multicolumn{10}{| l |}{Core intelligent approach}\\
\hline
1 hour periodicity	&	\multirow{3}{*}{00:00:38}	&	\multirow{3}{*}{00:00:0.38}	&	\multirow{3}{*}{00:00:0.38}	&	00:04:27	&	00:04:27	&	\multirow{3}{*}{05:00:00}	&	05:26:13	&	05:26:13	&	07:11:37\\
2 hour periodicity	&		&		&		&	00:05:37	&	00:05:37	&	&	05:16:22	&	05:16:22	&	06:22:34\\
4 hour periodicity	&		&		&		&	00:06:29	&	00:06:29	&	&	05:13:32	&	05:13:32	&	05:31:21\\
\hline
\end{tabular}

\end{sidewaystable}





\end{document}